\documentclass[10pt]{article}
\usepackage{bbm}
\usepackage[final]{graphics}
\usepackage{amsmath}
\usepackage{amsfonts,amsbsy}
\usepackage{amssymb}
\usepackage{color}
\usepackage{float}

\newcommand\bm\boldsymbol
\newcommand\ontop[2]{\genfrac{}{}{0pt}{}{#1}{#2}}

\newcommand{\be}{\begin{equation}}
\newcommand{\ee}{\end{equation}}
\newcommand{\ie}{\textsl{i.e.}}
\newcommand{\smeq}{\!\!\!=\!\!\!}
\newcommand{\f}{\frac}

\renewcommand{\d}{\ensuremath{\mathrm{d}}}

\newcommand{\bmpp}{\ensuremath{{\bm p}_\perp}}
\newcommand{\bmppa}[1]{\ensuremath{{\bm p}_{\perp{#1}}}}
\newcommand{\ppa}[1]{\ensuremath{p_{\perp{#1}}}}
\newcommand{\pz}{\ensuremath{p_z}}
\newcommand{\pza}[1]{\ensuremath{p_{z_{#1}}}}
\newcommand{\pp}{\ensuremath{p_\perp}}

\def\empile#1\over#2{\mathrel{\mathop{\kern 0pt#1}\limits_{#2}}}
\def\bs{\boldsymbol}

\def\TODO#1{}

\def\p{{\boldsymbol p}}
\def\P{{\boldsymbol P}}

\def\x{{\boldsymbol x}}

\newcommand{\slL}{\raise.15ex\hbox{$/$}\kern-.53em\hbox{$L$}}
\newcommand{\slP}{\raise.15ex\hbox{$/$}\kern-.53em\hbox{$P$}}
\newcommand{\slD}{\raise.15ex\hbox{$/$}\kern-.67em\hbox{$D$}}
\newcommand{\slp}{\raise.1ex\hbox{$/$}\kern-.63em\hbox{$p$}}
\newcommand{\slq}{\raise.1ex\hbox{$/$}\kern-.53em\hbox{$q$}}
\newcommand{\slv}{\raise.1ex\hbox{$/$}\kern-.63em\hbox{$v$}}
\newcommand{\slR}{\raise.15ex\hbox{$/$}\kern-.53em\hbox{$R$}}
\newcommand{\slQ}{\raise.15ex\hbox{$/$}\kern-.53em\hbox{$Q$}}
\newcommand{\slK}{\raise.15ex\hbox{$/$}\kern-.53em\hbox{$K$}}
\newcommand{\slk}{\raise.15ex\hbox{$/$}\kern-.53em\hbox{$k$}}
\newcommand{\slSigma}{\raise.15ex\hbox{$/$}\kern-.53em\hbox{$\Sigma$}}
\newcommand{\slcalP}{\raise.15ex\hbox{$/$}\kern-.63em\hbox{$\cal P$}}
\newcommand{\slcalA}{\raise.15ex\hbox{$/$}\kern-.63em\hbox{$\cal A$}}
\newcommand{\slA}{\raise.15ex\hbox{$/$}\kern-.73em\hbox{$A$}}
\newcommand{\slbfA}{\raise.15ex\hbox{$/$}\kern-.73em\hbox{${\imb A}$}}
\newcommand{\slpartial}{\raise.15ex\hbox{$/$}\kern-.53em\hbox{$\partial$}}
\newcommand{\sla}{\raise.15ex\hbox{$/$}\kern-.53em\hbox{$a$}}
\newcommand{\slb}{\raise.15ex\hbox{$/$}\kern-.53em\hbox{$b$}}
\newcommand{\slc}{\raise.15ex\hbox{$/$}\kern-.53em\hbox{$c$}}
\newcommand{\slC}{\raise.15ex\hbox{$/$}\kern-.63em\hbox{$C$}}
\newcommand{\sln}{\raise.15ex\hbox{$/$}\kern-.575em\hbox{$n$}}

\begin{document}

\title{\bf Kinetic theory of a longitudinally expanding system of
  scalar particles} \author{Thomas Epelbaum${}^1$, Fran\c cois
  Gelis${}^2$\\ Sangyong Jeon${}^1$, Guy Moore${}^{3}$, Bin Wu${}^2$}
\maketitle

\begin{itemize}
\item[{\bf 1}.] McGill University, Department of Physics\\ 3600 rue University, Montr\'eal QC H3A 2T8, Canada
\item[{\bf 2.}] Institut de physique th\'eorique, Universit\'e Paris Saclay\\
 CEA, CNRS, F-91191 Gif-sur-Yvette, France
\item[{\bf 3.}] Institut f\"ur Kernphysik, Technische Universit\"at Darmstadt\\ Schlossgartenstra{\ss}e 2, D-64289 Darmstadt, Germany
\end{itemize}

\begin{abstract}
  A simple kinematical argument suggests that the classical
  approximation may be inadequate to describe the evolution of a
  system with an anisotropic particle distribution. In order to verify
  this quantitatively, we study the Boltzmann equation for a
  longitudinally expanding system of scalar particles interacting with
  a $\phi^4$ coupling, that mimics the kinematics of a heavy ion
  collision at very high energy. We consider only elastic $2\to 2$
  scatterings, and we allow the formation of a Bose-Einstein condensate
  in overpopulated situations by solving the coupled equations for the
  particle distribution and the particle density in the zero mode. For
  generic CGC-like initial conditions with a large occupation number,
  the solutions of the full Boltzmann equation cease to display the
  classical attractor behavior sooner than expected; for moderate
  coupling, the solutions appear never to follow a classical attractor
  solution.
\end{abstract}

\section{Introduction and motivation}
\label{sec:intro}
\subsection{Context}
A long standing problem in the theoretical study of heavy ion
collisions is the time evolution of the pressure tensor and its
isotropization~\cite{BergeBSV1,BergeBSV3,BergeBSV2,BlaizGLMV1,DusliEGV2,EpelbG3,KurkeM3,KurkeM1,KurkeL1,HelleMST1,HelleMST2,WuR1}. This
question is closely related to the use of hydrodynamics in the
modeling of heavy ion collisions. Although the complete isotropy of
the pressure tensor is not necessary \cite{HelleJW1} for the validity
of hydrodynamical descriptions, the ratio of longitudinal to
transverse pressure, $P_{_L}/P_{_T}$ , should increase with time for a
smooth matching between the pre-hydro model and hydrodynamics.

In most models with boost invariant initial conditions, the
longitudinal pressure is negative immediately
after the collision \cite{FukusG1,ScheeRP1} (in the Color Glass
Condensate framework \cite{IancuLM3,IancuV1,Weige2,GelisIJV1,Gelis15},
this can be understood as a consequence of large longitudinal
chromo-electric and chromo-magnetic fields). On timescales of the
order of a few $Q_s^{-1}$ ($Q_s$ is the saturation momentum), the
longitudinal pressure rises, and reaches a positive value of
comparable magnitude to the transverse pressure. However, it is
unclear whether the scatterings are strong enough to sustain this mild
anisotropy, or whether the longitudinal expansion wins and causes the
anisotropy to increase.

This regime may be addressed by various tools. Indeed, it corresponds
to a period where the gluon occupation number is still large compared
to 1, but small compared to the inverse coupling $g^{ −2}$ so that a
quasi-particle picture may be valid. This means that the system could
in principle be described either in terms of fields or in terms of
particles. Since the occupation number is large, it is tempting to
treat the system as purely classical. In a description in terms of
fields, this amounts to considering classical solutions of the field
equations of motion, averaged over a Gaussian ensemble of initial
conditions whose variance is proportional to the initial occupation
number. In a description in terms of particles, \ie\ kinetic theory,
this amounts to keeping in the collision integral of the Boltzmann
equation only the terms that have the highest degree in the occupation
number \cite{MuellS1,Jeon3,MathiMT1}.

\subsection{Classical attractor scenario}
\label{sec:scen}
The field theory version of this classical approximation has been
implemented re\-cently \cite{BergeBSV4} for scalar fields with
longitudinal expansion, and it leads to a decrease of the ratio
$P_{_L}/P_{_T}$, like the power $\tau^{-2/3}$ of the proper time. This
behavior seems universal; it has been observed for a wide range of
initial conditions, and both for Yang-Mills theory and scalar field
theories with a point-like interaction (such as a $\phi^4$
interaction) \cite{BergeBSV1,BergeBSV3,BergeBSV2}. Based on this
observation, it was conjectured that the time evolution of
$P_{_L}/P_{_T}$ can be decomposed in three stages, nicely summarized
in Figure 3 of ref.~\cite{BergeSSV1}~:
\begin{itemize}
\item[{\bf i.}] A transient stage that depends on the details of the
  initial condition;
\item[{\bf ii.}] A universal scaling stage, called ``classical
  attractor,'' during which the dynamics is purely classical and the
  ratio $P_{_L}/P_{_T}$ decreases as  $\tau^{-2/3}$;
\item[{\bf iii.}] A final stage where the occupation number has become
  of order 1, and where quantum corrections are important. The
  isotropization of the pressure tensor may happen during this final
  stage.
\end{itemize}

\subsection{Classical approximation in anisotropic systems}
However, a possible difficulty with the classical approximation is
that the occupation number cannot be large uniformly at all momenta,
which may make this approximation unreliable since the dynamics
integrates over all the momentum modes. In particular, this may be the
case in anisotropic systems, as we shall explain now in a kinetic
theory framework. Suppose for discussion that the particle distribution has
become extremely anisotropic, with a narrow support in $\pz$,
\begin{align}
f(\p_\perp,\pz)\sim \delta(\pz)\,{\bf f}(\p_\perp)\,.\label{eq:initf}
\end{align}
This situation occurs in the pre-equilibrium stage of heavy ion
collisions, that can be described at leading order by a rapidity
independent classical color field.  In a non expanding
system\footnote{For a longitudinally expanding system, collisions will
  compete with the expansion, and the final outcome may depend on
  details of the scattering cross-section.}, we expect that $2\to 2$
scatterings will kick particles out of the transverse plane and
restore the isotropy of the distribution.
\begin{figure}[htbp]
\begin{center}
\resizebox*{4cm}{!}{\includegraphics{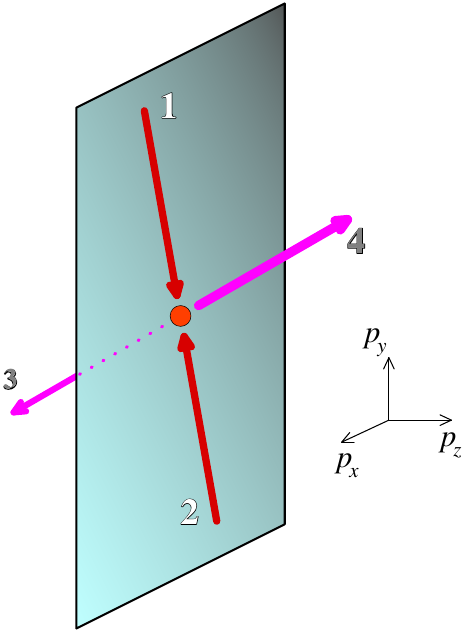}}
\end{center}
\caption{\label{fig:22scatt}Momentum-space picture of the $2\to 2$
  scattering contributing to isotropization. The particles 1 and 2 are
  in the transverse plane, while 3 and 4 have a non zero $\pz$.}
\end{figure}
The Boltzmann equation that describes the evolution of
$f(\p_\perp,p_z)$ should be able to capture this isotropization.  If
we track the momentum $p_4$ (see Figure \ref{fig:22scatt}), the
Boltzmann equation can be sketched as follows:
\begin{eqnarray}
\partial_t f_4
&\sim&
g^4\int\limits_{p_{1,2,3}}
\cdots
\big[\underline{f_1f_2(f_3+f_4)}-\underline{f_3f_4(f_1+f_2)}\big]
\nonumber\\
&&\qquad\qquad
+\,g^4\int\limits_{p_{1,2,3}}\cdots\big[f_1f_2-\underline{f_3f_4}\big]\,.
\label{eq:boltz}
\end{eqnarray}
We have separated the purely classical terms (cubic in $f$) from the
subleading terms that are quadratic in $f$. The classical
approximation amounts to dropping the quadratic terms. The usual
justification of this approximation is to say that the cubic terms are
much larger than the quadratic ones when $f\gg 1$. However, as we
shall see now, this counting is too naive when the support of $f$ is
anisotropic.

If $p_1$ and $p_2$ are in the transverse plane, then $\pza{3} + \pza{4} =
0$. Therefore, if we request that the particle 4 is produced outside
of the transverse plane, then we have $f_3 = f_4 = 0$. Because of
this, many pieces of the collision term (underlined in
eq.~(\ref{eq:boltz})) are zero. In particular, all the classical terms
vanish. The physical interpretation is clear: the $f^3$ terms
correspond to stimulated emission, which cannot happen when both final
particles lie in an empty region of phase space.

The only non-zero term is the one in $f_1f_2$, but one must go beyond
the classical approximation in order to capture it. This is true no
matter how large the distribution $f(\p)$ is inside its support,
\ie\ even if the classicality condition $f(\p)\gg 1$ is satisfied
there. Moreover, this argument can be trivially generalized to any
$n\to n'$ scattering process in kinetic theory\footnote{The classical
  approximation of a $n\to n'$ collision term contains only terms of
  degree $f^{n+n'-1}$. Because of longitudinal momentum conservation,
  if one of the momenta points out of the transverse plane, then there
  must be at least one out-of-plane momentum. Therefore, of the
  $n+n'-1$ distribution functions, at least one is zero.}. Therefore,
when the particle distribution is anisotropic, the classical
approximation artificially suppresses out-of-plane scatterings at
large angle, possibly resulting in wrong conclusions regarding
isotropization.

Now let us relax the $\delta$ function assumption, and consider the
case where the range of angles with large occupancy is finite but
narrow.  In particular, suppose that for momenta $p\sim Q$, the
particles mostly reside with $|p_z| < \delta Q$, where $\delta \ll 1$
describes how anisotropic the momentum distribution is.  We are
interested in scattering processes which move particles out of this
highly-occupied region.  So consider a $2\leftrightarrow 2$ scattering
process, where the final momentum $p_4$ lies outside this
highly-occupied region, so $f_4$ is small.  Within the classical
approximation, Eq.~(\ref{eq:boltz}) is dominated by the $f_1 f_2 f_3$
term.  But for all three of these occupancies to be large, we need
$|p_{1z}|,|p_{2z}|,|p_{3z}| < \delta Q$.  Since
$p_{z1} + p_{z2} = p_{z3} + p_{z4}$, this ensures that
$|p_{z4}| < 3 \delta Q$, that is, the final state particle produced in
a classical scattering must still have quite a small $|p_z|$ value.
We will refer to these as classical scatterings.
A scattering with $|p_{3z}|\sim Q$ and $|p_{z4}|\sim Q$ will always be
suppressed by a small occupancy in the classical approximation, so
these scatterings can be neglected classically, and only occur because
of the quantum $f_1 f_2$ term in Eq.~(\ref{eq:boltz}).  We will call
these quantum scatterings.

Now let us see whether scatterings with $|p_{z3}|,p_{z4}|\sim Q$ may still be
important, even if the occupancies $f(|p_z|\sim \delta Q) \gg 1$ are
large.  First of all, while the integrand in Eq.~(\ref{eq:boltz}) is
small for quantum scatterings, the integration phase space is much
larger for these processes.  Specifically, since $|p_{z1}|,|p_{z2}|
\sim \delta Q$ in all cases, and because of the momentum-conserving
delta function, the phase space for ``quantum'' scatterings is larger
than that for ``classical'' scatterings by a factor of $\sim 1/\delta$.
Therefore, even if the occupancy in the highly-occupied region is
$f\sim 1/\delta$, order-1 of the scatterings are ``quantum.''
That is, order-1 of the scatterings involve quantum effects when the
\textsl{angle-averaged} occupancy below $p\sim Q$ is order 1.
This argument may not apply in gauge theories, where the matrix
element is strongly enhanced for small-angle processes, but it should
be valid in a scalar theory where the matrix element is isotropic.

In addition, for many purposes, an individual scattering resulting in
$|p_{z4}| \sim Q$ is much more important than a scattering resulting
in $|p_{z4}| \sim 3 \delta Q$.  A particle's contribution to the
longitudinal pressure involves $p_z^2 / p^2$.  The particle produced
in a ``quantum'' scattering, with $|p_z|\sim Q$, contributes more
to the longitudinal pressure than the classically-scattered
$|p_z|\sim \delta Q$ particle, by a factor of $\sim 1/\delta^2$.
Therefore, in terms of generating longitudinal pressure, each
``quantum'' scattering contributes a more important effect, by a
factor of $1/\delta^2$.  Combining this factor with the larger phase
space for quantum scattering, we find that, for the purposes of
understanding the longitudinal pressure, the classical approximation
already starts to fail when $f(p_z\sim 0,|p|\sim Q) \sim \delta^{-3}$,
which is when the angle-averaged occupancy is $1/\delta^2 \gg 1$.

\begin{figure}[htbp]
\begin{center}
\resizebox*{10cm}{!}{\includegraphics{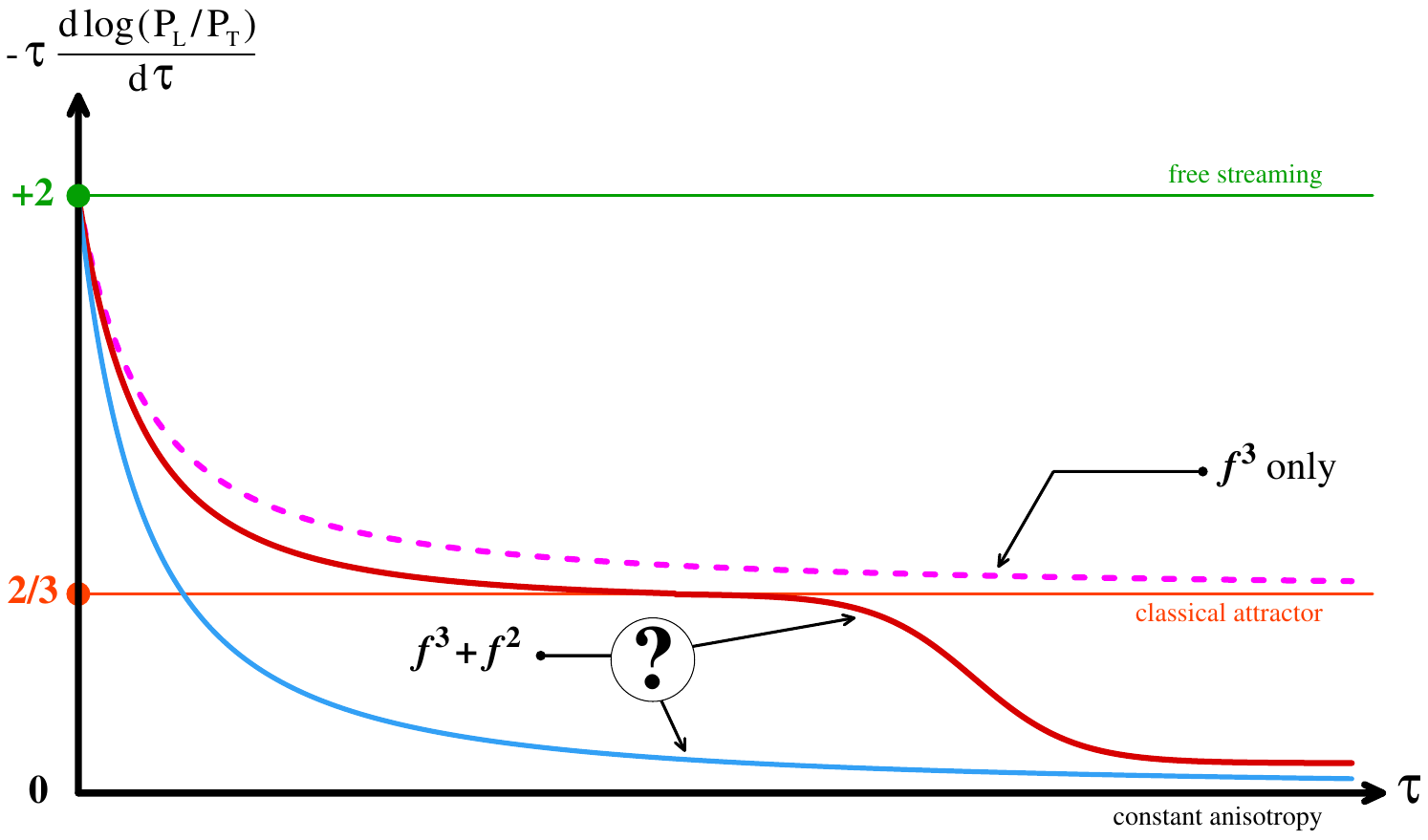}}
\end{center}
\caption{\label{fig:iso}Examples of possible behaviors of the logarithmic
  derivative of the ratio $P_{_L}/P_{_T}$. Dotted curve: classical
  approximation where one keeps only the $f^3$ terms. Red curve: full
  collision term, if there exists a classical attractor. Blue
  curve: full collision term, if there is no classical attractor.}
\end{figure}

This suggests that the classical approximation may lead to missing
some contributions that are important for isotropization in theories
where the cross-section is dominated by large-angle scatterings, such
as a $\phi^4$ scalar theory. When the classical approximation is used
in this theory, it has been observed in ref.~\cite{BergeBSV4} that the
ratio $P_{_L}/P_{_T}$ decreases like $\tau^{-2/3}$ at late times for a
longitudinally expanding system (dotted curve in Figure
\ref{fig:iso}). Given the above argument, the $f^2$ terms that are
neglected in the classical approximation could alter this
behavior sooner than one might naively expect,
ending the growth of anisotropy and giving a ratio
$P_{_L}/P_{_T}\sim \tau^0$ at late times. If the $f^2$ terms are truly
negligible over some extended period of time, then the full Boltzmann
equation should lead to the red curve in Figure \ref{fig:iso},
in which the system spends some time stuck into a classical attractor
before eventually leaving it in order to isotropize (this red curve
corresponds to the 3-stage scenario of Section \ref{sec:scen}). In
contrast, if the $f^2$ terms are important from the start, one may
get the behavior illustrated by the blue curve, where the power law
$\tau^{-2/3}$ does not play any particular role in the evolution of
$P_{_L}/P_{_T}$ and a classical attractor would not exist.

\subsection{Contents}
In the rest of this paper, in order to assess quantitatively this
issue, we solve the Boltzmann equation for an expanding system of
scalar bosons with a $\phi^4$ interaction, both for the full collision
kernel and its classical approximation.  In Section
\ref{sec:boltz}, we discuss the Boltzmann equation for a
longitudinally expanding system and prepare the stage for its
numerical resolution. We describe our algorithm in Section
\ref{sec:num}, and the numerical results are exposed in Section
\ref{sec:Resolution of the Boltzmann equation}. Section
\ref{sec:concl} is devoted to a summary and concluding remarks. Some
more technical material and digressions are presented in
appendices. In appendix \ref{sec:Non-expanding system}, we present
results on the isotropization of the pressure tensor in a
non-expanding system, that corroborate the fact that the $f^2$ terms
play an essential role. In appendix \ref{sec:apa}, we derive an
analytical expression for the azimuthal integrals of the $2\to 2$
collision term.  Additional details about our algorithm can be found
in appendices \ref{sec:apb} and \ref{sec:apc}. In the appendix
\ref{app:dsmc}, we present an alternate algorithm for solving the
Boltzmann equation, based on the direct simulation Monte-Carlo (DSMC)
method.

\section{Boltzmann equation for expanding systems}
\label{sec:boltz}
\subsection{Notation}
\label{sec:Notations}
In the following, we consider partially anisotropic particle
distributions that have a residual axial symmetry around the $z$ axis,
\begin{equation}
f(\bmpp,\pz)=f(\pp,\pz)
\end{equation}
($p_\perp\equiv \big|\p_\perp\big|$). For simplicity, we do not write
explicitly the space and time dependence of $f$. The function $f$
depends smoothly on momentum, except possibly at $p_z=p_\perp=0$ if
there is a Bose-Einstein condensate (see Section
\ref{sec:Bose-Einstein Condensate}).  We also assume that $f$ is even
in the longitudinal momentum $p_z$
\begin{equation}
f(\pp,-\pz)=f(\pp,\pz)\,.\label{eq:evenlongassump}
\end{equation}
The Lorentz covariant form of the Boltzmann equation reads
\begin{equation}
  (p^\mu\partial_\mu)\:f(p) = \omega_\p\: C_\p[f]\,,
\label{eq:bolt-SK}
\end{equation}
where $C_\p[f]$ is the collision term (see the subsection \ref{sec:
  Collision term}).  This form of the equation can then be specialized
to any system of coordinates.  We will focus on the collision term
which arises at lowest order in the coupling, which for $\phi^4$
theory is elastic $2 \leftrightarrow 2$ scattering.

For a system that expands in a boost invariant way in the longitudinal
direction, the most appropriate system of coordinates is
$(\tau,\eta,\x_\perp)$ for the space-time coordinate and
$(y,\p_\perp)$ for the momentum of the particle, which are related to
the usual Cartesian coordinates and momenta by
\begin{eqnarray}
&&\tau\equiv \sqrt{t^2-z^2}
\quad,\quad
\eta\equiv \frac{1}{2}\ln\left(\frac{t+z}{t-z}\right)
\quad,\quad
y\equiv\frac{1}{2}\ln\left(\frac{p_0+p_z}{p_0-p_z}\right)
\nonumber\\
&&\x_\perp\equiv(x,y)\quad,\quad\p_\perp\equiv(p_x,p_y)\,.
\label{eq:coord}
\end{eqnarray}
The assumed boost invariance of the problem implies that the
distribution $f$ does not depend separately on $\eta$ and $y$, but
only on the difference $y-\eta$. Therefore, it is sufficient to derive
the equation for the distribution at mid-rapidity, $\eta=0$.  For
simplicity, we will further assume that the distribution is
independent of the transverse position.

\subsection{Free streaming term}
\label{sec:Frstr}
In the system of coordinates defined in eqs.~(\ref{eq:coord}), the
left-hand side of the Boltzmann equation can be rewritten as
\begin{equation}
(p^\mu\partial_\mu)\: f= p^\tau \partial_\tau\: f+\frac{p^\eta}{\tau} \partial_\eta\: f\,,
\end{equation}
where we have defined
\begin{equation}
p^\tau\equiv M_\p\cosh(y-\eta)\quad,\quad
p^\eta\equiv M_\p\sinh(y-\eta)\quad,\quad M_\p\equiv \sqrt{p_\perp^2+m^2}\,.
\end{equation}
In a boost invariant system, $f$ depends on $y$ and $\eta$ only via
the difference $y-\eta$ and it is sufficient to consider only the
distribution at mid-rapidity, $\eta=0$, for which we have
\begin{equation}
(p^\mu\partial_\mu)\; f=
\left[M_\p\cosh(y)\;\partial_\tau-\frac{M_\p\sinh (y)}{\tau}\;\partial_y\right] f\,.
\end{equation}
This formula implicitly assumes that the distribution $f$ is given in
terms of the transverse momentum $\p_\perp$ and the rapidity $y$.
Energy and momentum conservation will take a simpler form if we
express it in terms of the energy $\omega_p$ and longitudinal momentum
$p_z$, instead of $\p_\perp$ and $y$.  Since $\pz=M_\p \sinh (y)$ and
$\omega_p=M_\p \cosh (y)$, one gets\footnote{The more familiar form
\begin{equation*}
(p^\mu\partial_\mu)\;f=
\omega_{p}\left[\partial_\tau-\frac{\pz}{\tau}\partial_{\pz}\right]\, f(p_\perp,p_z)
\end{equation*}
is obtained when one expresses $f$ in terms of $\pz$ and $\p_\perp$.}
\begin{equation}
(p^\mu\partial_\mu)\; f=
\omega_{p}\left[\partial_\tau-\frac{\pz^2}{\tau \omega_p}\partial_{\omega_p}-\frac{\pz}{\tau}\partial_{\pz}\right]\, f(\omega_p,\pz)\,.\label{eq:Dopexpand}
\end{equation}

\subsection{Collision term}
\label{sec: Collision term}
If we consider only $2\to 2$ scatterings, the right-hand side of the
Boltzmann equation contains integrals over the on-shell phase spaces
of three particles. This 9-dimensional integral can be reduced to a
5-dimensional integral by using energy and momentum conservation. A
further simplification results from our assumption that the
distribution is invariant under rotations around the $p_z$ axis. The
Boltzmann equation reads
\begin{equation}
\left[\partial_\tau-\frac{\pza{1}^2}{\tau \omega_{p_1}}\partial_{\omega_{p_1}}-\frac{\pza{1}}{\tau}\partial_{\pza{1}}\right]\, f(\omega_{p_1},\pza{1})
=C_{p_1}[f]\label{eq:bolt}
\end{equation}
with
\begin{equation}
C_{p_1}[f]\equiv
\frac{g^4}{4\omega_{p_1}}\int\limits_{\p_{2,3,4}}
(2\pi)^4\delta^{(4)}(P_1+P_2-P_3-P_4)\:F_{_{\rm nc}}(\{P_i\})\,.\label{eq:firstcollnc}
\end{equation}
In this equation, $\int_\p$ denotes the integration over the invariant
phase-space
\begin{equation}
\int_\p\quad\equiv\quad\int\frac{\d^3\p}{(2\pi)^32\omega_p}\: ,
\end{equation}
and $F_{_{\rm nc}}(\{P_i\})$ is the factor that contains the particle
distribution\footnote{The subscript ``nc'' means ``no condensate''. In
  the next subsection, we will generalize this equation to the case
  where  Bose-Einstein condensation can happen.},
\begin{equation}
F_{_{\rm nc}}(\{P_i\})
\equiv
f_3f_4(1+f_1)(1+f_2)-f_1f_2(1+f_3)(1+f_4)\: .
\label{eq:Fnc}
\end{equation}
(We use the abbreviation $f_i\equiv f(p_i)$.)

The invariance under rotation around the $p_z$ axis can be used in order
to perform analytically all the integrals over azimuthal angles (we
use the same procedure as in the case of an $O(3)$ rotational
invariance, see refs.~\cite{SemikT1,SemikT2}). In order to
achieve this, let us first write
\begin{equation}
(2\pi)^2\,\delta(\bmppa{1}+\bmppa{2}-\bmppa{3}-\bmppa{4})=
\int \d^2 {\bm x}_\perp\: e^{i\,{\bm x}_\perp\cdot(\bmppa{1}+\bmppa{2}-\bmppa{3}-\bmppa{4})}\,.
\end{equation}
By combining this with the integrations over the transverse momenta,
we get
\begin{eqnarray}
&&
\int
\frac{\d^2 \bmppa{2}}{(2\pi)^22\omega_{p_2}}
\frac{\d^2 \bmppa{3}}{(2\pi)^22\omega_{p_3}}
\frac{\d^2 \bmppa{4}}{(2\pi)^22\omega_{p_4}}\:
(2\pi)^2 \delta^{(2)}(\bmppa{1}+\bmppa{2}-\bmppa{3}-\bmppa{4})
 =\nonumber\\
&&\qquad=\frac{1}{32\pi^2}\int\limits_m^{+\infty}\d \omega_{p_2}\,\d \omega_{p_3}\,\d \omega_{p_4}
\int\limits_0^{+\infty} \d x_\perp \:x_\perp\:
\prod_{i=1}^4J_0(p_{\perp i}x_\perp)\,.
\end{eqnarray}
The integral over $x_\perp$ depends only on the transverse momenta
$\{\ppa{i}\}$, but not on the distribution function $f$. It can
therefore be calculated once for all. In the appendix \ref{sec:apa},
we obtain an explicit formula for this integral in terms of the
Legendre elliptic $K$ function. Defining
\begin{eqnarray}
r_1&\equiv&\max((\ppa{1}-\ppa{2})^2,(\ppa{3}-\ppa{4})^2)\nonumber\\
r_2&\equiv&\min((\ppa{1}+\ppa{2})^2,(\ppa{3}+\ppa{4})^2)\nonumber\\
r_3&\equiv&\min((\ppa{1}-\ppa{2})^2,(\ppa{3}-\ppa{4})^2)\nonumber\\
r_4&\equiv&\max((\ppa{1}+\ppa{2})^2,(\ppa{3}+\ppa{4})^2)\,,
\end{eqnarray}
we obtain
\begin{equation}
{\bs I}_4(\{p_{\perp i}\})
\equiv \int_0^{+\infty} \d x_\perp \: x_\perp\:
\prod_{i=1}^4J_0(p_{\perp i}x_\perp)
=
\frac{4\,K\left({\frac{(r_2-r_1)(r_4-r_3)}{(r_4-r_1)(r_2-r_3)}}\right)}{\pi^2\sqrt{(r_4-r_1)(r_2-r_3)}}
\end{equation}
where\footnote{In the appendix \ref{sec:apa}, we present a simple and
  fast algorithm for evaluating $K(z)$.}
\begin{equation}
{K}(z)\equiv\int_0^{\pi/2}\frac{\d\theta}{\sqrt{1-z\sin^2\theta}}\, .
\end{equation}
In terms of this integral, the collision term reads
\begin{eqnarray}
&&C_{p_1}[f]=
\frac{g^4}{256\pi^3\,\omega_{p_1}}
\int\limits_m^{+\infty}\!
\d \omega_{p_2}\,\d \omega_{p_3}\,\d \omega_{p_4}\;{\bs I}_4(\{p_{\perp i}\})
\int
\d \pza{2}\,
\d \pza{3}\,
\d \pza{4}\nonumber\\
&&\times
\delta(\omega_{p_1}+\omega_{p_2}-\omega_{p_3}-\omega_{p_4})\delta(\pza{1}+\pza{2}-\pza{3}-\pza{4})
\;F_{_{\rm nc}}(\{P_i\})\,.
\label{eq:boltbefdelta}
\end{eqnarray}
For the sake of brevity, we have not written explicitly the boundaries
of the integration range on $p_{z_2}, p_{z_3}, p_{z_4}$. They are given by
\begin{equation}
-\sqrt{\omega_{p i}^2-m^2}\le p_{z i}\le  \sqrt{\omega_{p i}^2-m^2}\, .
\end{equation}
Eq (\ref{eq:boltbefdelta}) reduces to a 4-dimensional integral after
taking into account the two delta functions, which is doable
numerically. Assuming that we encode the momenta with $(\omega_p,\pz)$
there is no need to perform any interpolation when using the delta
functions to eliminate two of the integration variables, which is
useful for fulfilling with high accuracy the conservation of energy
and particle number. After this reduction that eliminates the
variables $\omega_{p_2}$ and $p_{z_2}$, eq.~(\ref{eq:boltbefdelta})
reduces to
\begin{equation}
C_{p_1}[f]=\frac{g^4}{256\pi^3\,  \omega_{p_1}}
\int_{\mathcal{C}_{\omega}}\d \omega_{p_3}\,\d \omega_{p_4}\,
\int_{\mathcal{C}_{\pz}}\,\d \pza{3}\,\d \pza{4}
\;{\bs I}_4(\{p_{\perp i}\})\;F_{_{\rm nc}}(\{P_i\})\,.\label{eq:boltafdelta}
\end{equation}
At each time step, $C_{p_1}[f]$ must be calculated for each value of
$\pza{1}$ and $\omega_{p_1}$. Therefore, if we discretize the
variables $\omega$ and $\pz$ with $N_{\rm f}$ and $N_z$ lattice points
respectively, the computational cost for each time step scales as
$(N_{\rm f}N_z)^3$.

\begin{figure}[htbp]
\begin{center}
\resizebox*{!}{9cm}{\includegraphics{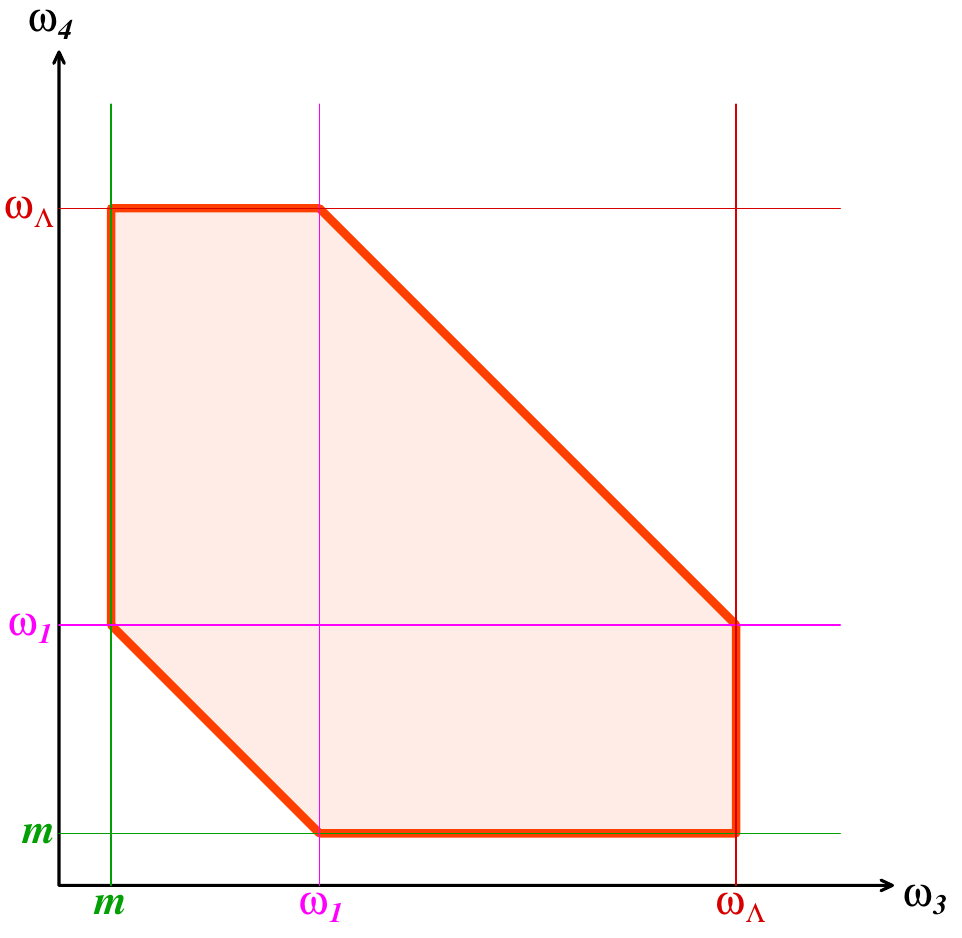}}
\end{center}
\caption{\label{fig:om34}Integration domain for the variables $\omega_{p_3}$ and $\omega_{p_4}$.}
\end{figure}

\begin{figure}[htbp]
\begin{center}
\resizebox*{!}{12cm}{\includegraphics{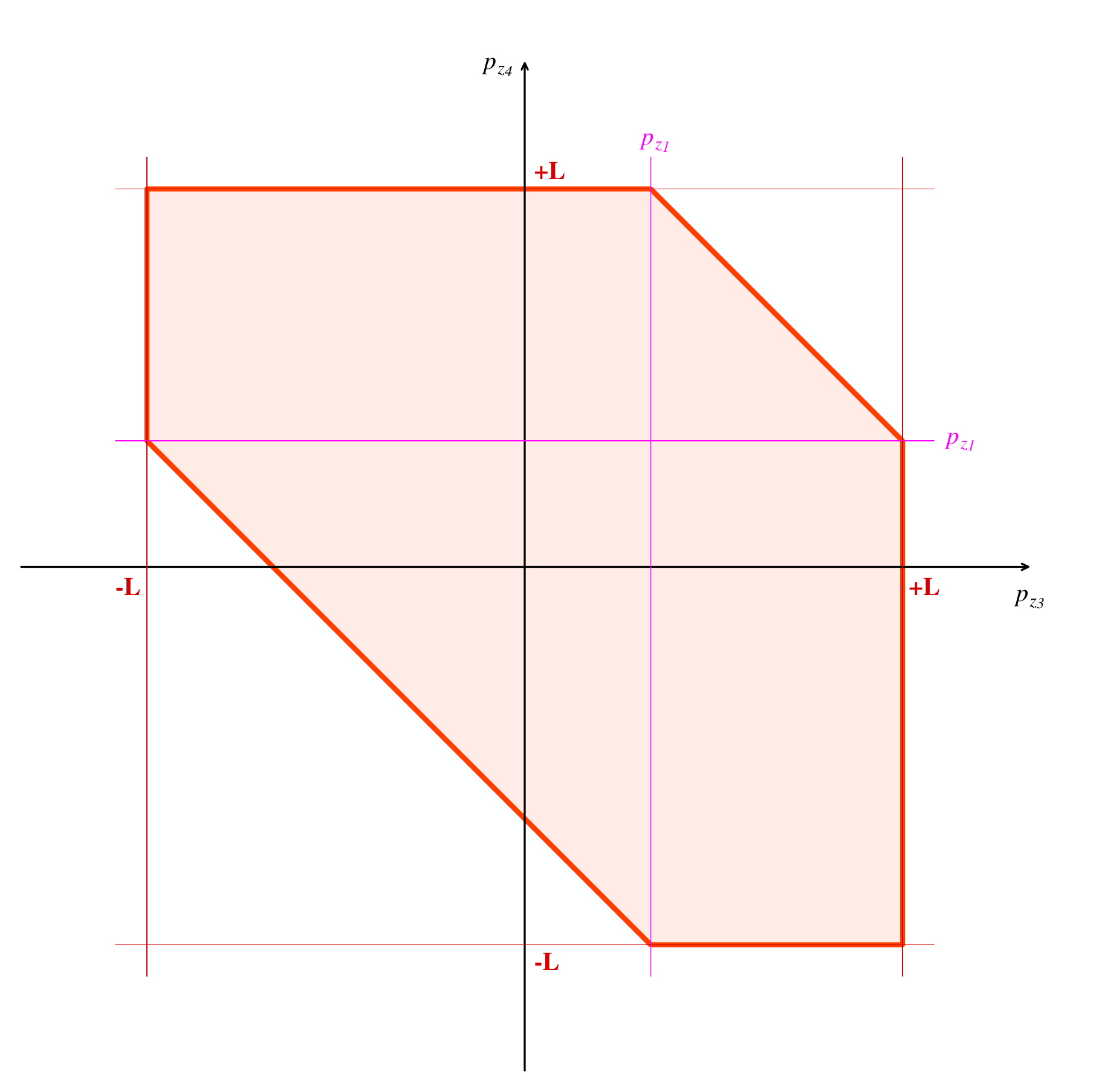}}
\end{center}
\caption{\label{fig:pz34}Integration domain for the variables $\pza{3}$ and $\pza{4}$.}
\end{figure}

In the numerical implementation, the allowed energy and momentum ranges
must be bounded.  We will denote the maximum allowed energy as
$\omega_{_{\Lambda}}$ and the maximum allowed longitudinal momentum as
$L$, so that
\begin{equation}
m\le\omega_p\le\omega_{_{\Lambda}}\quad,\qquad -L\le p_z \le +L\, .
\label{eq:bounds}
\end{equation}
The integration domain $\mathcal{C}_{\omega}$ for $\omega_{p_3}$ and
$\omega_{p_4}$ is the bounded domain shown in Figure
\ref{fig:om34}.
Likewise, if we assume that $\pza{1}>0$ (since $f$ is even in $\pz$),
the integration domain $\mathcal{C}_{\pz}$ for $\pza{3}$ and $\pza{4}$
is the domain shown in Figure \ref{fig:pz34}.
The peculiar shape of these domains comes from the fact that
$\omega_{p_2}$ and $\pza{2}$ extracted from the delta functions must
themselves have values which lie within the bounds
(\ref{eq:bounds}). The variable $\pz$ can be positive or negative, and
although $f$ is even in $\pz$, we must integrate over both
$\pza{3,4}\ge 0$ and $\pza{3,4}\le 0$ inside the collision term,
because of the asymmetric shape of the integration domain. Note that
there is also an extra constraint (not represented on these diagrams
because it relates $\omega$ and $\pz$) that must be satisfied,
\begin{equation}
m^2+\pz^2\le \omega_p^2\, ,
\end{equation}
for the transverse momentum to be defined.

\subsection{Bose-Einstein condensation}
\label{sec:Bose-Einstein Condensate}
If only elastic collisions are taken into account, a Bose-Einstein
condensate (BEC) may appear in the system if the initial condition is
overpopulated \cite{BlaizGLMV1,EpelbG1,BlaizLM2,BlaizWY1,BergeS4}. When this
happens, the particle distribution has a singularity at zero momentum,
that we can represent by\footnote{Our convention for the integral of a
  delta function over the positive real semi-axis is
\begin{equation*}
  \int_0^{+\infty}\d x\;\delta(x)=\tfrac{1}{2}\, ,
\end{equation*}
so that the integral of the second term with the measure
$d^3\p/(2\pi)^3$ gives $n_c$.}
\begin{align}
f(\omega_{p_1},\pza{1})
\quad\rightarrow\quad
f(\omega_{p_1},\pza{1})+\frac{8\pi^2}{m} \delta(\omega_{p_1}-m)\delta(\pza{1})\,n_c\,.
\label{eq:singcond}
\end{align}
After this redefinition, $f$ describes the smooth part of the
occupation number and $n_c$ is the particle density in the condensate.
The Boltzmann equation (\ref{eq:bolt}) needs to be revised to account
for $n_c$. Indeed, one can now have particle-condensate
interactions. One can easily check that the $2\leftrightarrow 2$
processes cannot involve more than one particle from the
condensate. The modified Boltzmann equation thus reads
\begin{equation}
\left[\partial_\tau-\frac{\pza{1}^2}{\tau \omega_{p_1}}\partial_{\omega_{p_1}}
-\frac{\pza{1}}{\tau}\partial_{\pza{1}}\right]\, f(\omega_{p_1},\pza{1})=
C_{p_1}[f]
+
C^{_{1c\leftrightarrow 34}}_{p_1}[f]
+
C^{_{12\leftrightarrow c4}}_{p_1}[f]
\label{eq:Bfc}
\end{equation}
where $C^{_{1c\leftrightarrow 34}}_{p_1}[f]$ is the contribution from
collisions between the particle of momentum $p_1$ that we are tracking
and a particle from the condensate,
\begin{equation}
\p_1+{\bs 0}_c\quad\longleftrightarrow\quad \p_3+\p_4\, ,
\end{equation}
while $C^{_{12\leftrightarrow c4}}_{p_1}[f]$ stands for a collision
between the particle $p_1$ that we are tracking and another particle
of momentum $p_2$ to give a final state with a particle in the
condensate and a particle of momentum $p_4$,
\begin{equation}
\p_1+\p_2\quad\longleftrightarrow\quad {\bs 0}_c+\p_4\, .
\end{equation}
This term should be doubled, to account for the fact that the
condensate particle can be the particle $3$ or the particle $4$.
Following the same procedure\footnote{Here one can directly use the
  formula (\ref{eq:int3bessel}) for the integral of three Bessel
  functions.}  as in Section \ref{sec: Collision term}, we find
\begin{eqnarray}
&&
C^{_{1c\leftrightarrow 34}}_{p_1}[f]
=
\frac{g^4}{128\pi^2}\frac{n_c}{m\omega_{p_1}}
\!\int\! \d{p}_{z_4}\d\omega_{p_4}
\left[\frac{f_3f_4\!-\!f_1(1\!+\!f_3\!+\!f_4)}{{\cal A}(p_{\perp1},p_{\perp3},p_{\perp4})}
\right]_{\ontop{\omega_3=m+\omega_1-\omega_4}{{p}_{z_3}={p}_{z1}-{p}_{z_4}}}\nonumber\\
&&
C^{_{12\leftrightarrow c4}}_{p_1}[f]
=
\frac{g^4}{64\pi^2}\frac{n_c}{m\omega_1}
\!\int\! \d{p}_{z_4}\d\omega_{p_4}
\left[
\frac{f_4(1\!+\!f_1\!+\!f_2)\!-\!f_1f_2}{{\cal A}(p_{\perp1},p_{\perp2},p_{\perp4})}
\right]_{\ontop{\omega_2=m+\omega_4-\omega_1}{{p}_{z_2}={p}_{z_4}-{p}_{z1}}}\, ,
\nonumber\\
&&
\label{eq:Cc34}
\end{eqnarray}
where $\mathcal{A}(x,y,z)$ is the area of the triangle of edges $x,y,z$~:
\begin{align}
\mathcal{A}(x,y,z)\equiv\frac{1}{4} \sqrt{(x+y+z)(x+y-z)(x-y+z)(-x+y+z)}\, .
\end{align}
Note that in these collision terms, the 2-dimensional integration
domains of Figures \ref{fig:om34} and \ref{fig:pz34} collapse to
1-dimensional domains (see the appendix \ref{sec:apb}).

The equation for the evolution of the particle density $n_c$ in the
condensate can be obtained from eq.~(\ref{eq:boltafdelta}), by
replacing the particle $p_1$ that we are tracking with the singular
part of eq.~(\ref{eq:singcond}), and then integrating over $p_1$. By
doing so we obtain the following equation for the condensate
\begin{equation}
\tau^{-1}\partial_\tau (\tau n_c)
=
\frac{g^4}{512\pi^4}\frac{n_c}{m}
\int\limits_{\mathcal{C}_{\omega}}\d \omega_{p_3}\,\d \omega_{p_4}\,
\int\limits_{\mathcal{C}_{\pz}}\,\d \pza{3}\,\d \pza{4}
\;\frac{f_3f_4-f_2(1+f_3+f_4)}{{\cal A}(p_{\perp2},p_{\perp3},p_{\perp4})}\, ,
\label{eq:Cc1}
\end{equation}
where the integration domains are those of Figures \ref{fig:om34}
and \ref{fig:pz34} (with $\omega_{p_1}=m$ and $p_{z_1}=0$).

\subsection{Conservation laws}
\label{sec:cons}
The Boltzmann equation fulfills several conservation laws, that play
an important role in determining the form of the equilibrium particle
distribution and also in assessing the accuracy of algorithms employed
for solving the equation numerically. Each collision conserves energy
and momentum. In addition, since we are only considering elastic
scatterings in this paper, the collisions also conserve the number of
particles.

The particle density is given by
\begin{equation}
n=
n_c+
\underbrace{\frac{1}{4\pi^2}\int\omega_p\,\d \omega_p
\d \pz\;f(\pp,\pz)}_{n_{\rm nc}}\,.
\end{equation}
Note that this is a density of particles per unit of rapidity
$\eta$. Since a given interval of $\eta$ corresponds to a volume that
expands linearly with the proper time $\tau$, the conservation of the
total number of particles implies that
\begin{equation}
\tau\,n=\mbox{constant}\,.
\label{eq:n-cons}
\end{equation}

Likewise, the components of the energy-momentum tensor are given by
\begin{equation}
T^{\mu\nu}=
\delta^{\mu 0}\,\delta^{\nu 0}\, m\, n_c
+
\underbrace{\frac{1}{4\pi^2}\int \d \omega_p
\d \pz\; p^\mu p^\nu\; f(\pp,\pz)}_{T^{\mu\nu}_{\rm nc}}
\, .
\label{eq:Tmunu}
\end{equation}
($p^\mu\equiv (\omega_p,\pp,\pz)$.) In a longitudinally expanding system,
the conservation of energy and momentum, $\partial_\mu T^{\mu\nu}=0$,
becomes
\begin{equation}
\partial_\tau \epsilon+\frac{\epsilon+P_{_L}}{\tau}=0\, ,
\label{eq:T-cons}
\end{equation}
where $\epsilon\equiv T^{00}$ is the energy density and $P_{_L}\equiv
T^{33}$ is the longitudinal pressure.

The fact that the solutions of the Boltzmann equation satisfy the
conservation equations (\ref{eq:n-cons}) and (\ref{eq:T-cons}) is a
consequence of the delta function
$\delta(\omega_{p_1}+\omega_{p_2}-\omega_{p_3}-\omega_{p_4})$ and of
the symmetries of the collision term under various exchanges of the
particles $1,2,3,4$.  Namely, the integrand in the collision term is
symmetric under the exchange of the initial state or final state
particles~:
\begin{equation}
P_1\quad\longleftrightarrow\quad P_2\quad,\qquad
P_3\quad\longleftrightarrow\quad P_4\, ,
\label{eq:sym1}
\end{equation}
and antisymmetric if we swap the initial and final states:
\begin{equation}
(P_1,P_2)\quad\longleftrightarrow\quad(P_3,P_4)\, .
\label{eq:sym2}
\end{equation}
Therefore, any approximation scheme used in a numerical algorithm
should aim at satisfying these properties with high accuracy. The
easiest way to implement the delta functions without loss of accuracy
is to use a lattice with a constant spacing in the variables
$\omega_p$ and $p_z$. By doing this, one is guaranteed that the values
of $\omega_\p$ and $p_z$ obtained by solving the constraints provided
by the delta functions are also points on this grid. In addition, the
quadrature formulas used for approximating the integrals in the
collision term should lead to
\begin{eqnarray}
&&
\int \omega_{p_1}\,\d\omega_{p_1}\d p_{z_1}\;C_{p_1}[f]=0\,,
\nonumber\\
&&
\int \omega_{p_1}^2\,\d\omega_{p_1}\d p_{z_1}\;C_{p_1}[f]=0\,,
\nonumber\\
&&
\int \omega_{p_1}p_{z_1}\,\d\omega_{p_1}\d p_{z_1}\;C_{p_1}[f]=0\, ,
\label{eq:cons1}
\end{eqnarray}
which are consequences of the symmetries~(\ref{eq:sym1}) and
(\ref{eq:sym2}). In other words, even if these symmetries are
manifest in the collision kernel, one should be careful
not to violate them with an improper choice of the quadrature
weights. As we will see, our numerical scheme respects these
symmetries, so that the conservation laws can be satisfied up to
machine precision.

It is also important to realize the role played in the
conservation laws by the terms
\begin{equation}
-\left[\frac{\pza{1}^2}{\tau \omega_{p_1}}\partial_{\omega_{p_1}}
+\frac{\pza{1}}{\tau}\partial_{\pza{1}}\right]\, f(\omega_{p_1},\pza{1})
\end{equation}
that appear on the left-hand side of the Boltzmann equation. For
instance, these terms provide the term $(\epsilon+P_{_L})/\tau$ in
eq.~(\ref{eq:T-cons}), since we have
\begin{equation}
-\frac{1}{4\pi^2}\int \d\omega_{p_1}\d p_{z_1}\,\omega_{p_1}^2\;
\left[\frac{\pza{1}^2}{\tau \omega_{p_1}}\partial_{\omega_{p_1}}
+\frac{\pza{1}}{\tau}\partial_{\pza{1}}\right]\, f(\omega_{p_1},\pza{1})
=\frac{\epsilon+P_{_L}}{\tau}\, .
\end{equation}
However, this identity relies on a cancellation between the boundary
terms that result from the integration by parts on $\omega_{p_1}$ and
$p_{z_1}$. This must be kept in mind when discretizing the free
streaming part of the Boltzmann equation, in order to avoid
introducing violations of the conservation laws through these boundary
terms.

\subsection{Classical approximation}
\label{sec:class-approx}
A central question in this paper is the interplay between the
classical approximation and isotropization. Each computation will
therefore be performed twice:
\begin{itemize}
\item[{\bf i.}] With the full expression for the combination of
  distribution functions that appear in the equations (\ref{eq:Fnc}),
  (\ref{eq:Cc34}) and (\ref{eq:Cc1});
\item[{\bf ii.}] In the classical approximation where only the cubic
  terms in the particle distribution are kept. This entails the
  following changes:
  \begin{align}
    &\mbox{Eq.~(\ref{eq:Fnc}):}&\mbox{r.h.s.}&&\longrightarrow&&f_3f_4(f_1\!+\!f_2)-f_1f_2(f_3\!+\!f_4)\nonumber\\
    &\mbox{Eq.~(\ref{eq:Cc34}):}  &f_3f_4\!-\!f_1(1\!+\!f_3\!+\!f_4)&&\longrightarrow&&
f_3f_4\!-\!f_1(f_3\!+\!f_4)\nonumber\\
    &&f_4(1\!+\!f_1\!+\!f_2)\!-\!f_1f_2&&\longrightarrow&&
f_4(f_1\!+\!f_2)\!-\!f_1f_2\nonumber\\
    &\mbox{Eq.~(\ref{eq:Cc1}):}  &f_3f_4-f_2(1\!+\!f_3\!+\!f_4)&&\longrightarrow&&
f_3f_4-f_2(f_3\!+\!f_4)\nonumber\\
&&&&&&
\label{eq:classapprox}
  \end{align}
\end{itemize}
As will become clear in the description of our algorithm in the next
section, we use fixed cutoffs on the energy
$\omega_\p\le\omega_{_{_\Lambda}}$ and on the longitudinal momentum
$|p_z|\le L$. These cutoffs are not exactly the same as in the
implementation of the classical approximation in classical lattice
field theory, where one uses fixed cutoffs on the transverse momentum
and on the Fourier conjugate $\nu$ of the rapidity. Indeed,
$\nu\approx p_z\tau$, so that a fixed cutoff on $p_z$ roughly
corresponds to a cutoff on $\nu$ that grows linearly with
time.  Fortunately, we do not expect physical effects from these
cutoffs if they are taken high enough, since the occupancy typically
falls off exponentially at large energy and momentum.

Note that there is a variant of the classical approximation defined
in~(\ref{eq:classapprox}), in which each distribution function $f$ is
replaced by $f+\tfrac{1}{2}$. It is well known that this \textit{Ansatz}
provides the correct quadratic terms \cite{MuellS1,Jeon3}, accompanied
by some spurious terms that are linear in the distribution
function. This variant is known to suffer from a severe ultraviolet
cutoff dependence, when the cutoff becomes large compared to the
physical scales \cite{EpelbGTW1} (this property is closely related to
the non-renormalizability of a variant of the classical approximation
in quantum field theory, where one includes the zero point vacuum
fluctuations \cite{EpelbGW1}). For this reason, our algorithm for the
Boltzmann equation in a longitudinally expanding system cannot be
employed to study this alternate classical approximation, because of
its fixed cutoff in $p_z$, while the physical $p_z$s decrease with
time due to the expansion. In appendix \ref{sec:Non-expanding
  system}, where we consider the question of isotropization in a
non-expanding system, we have also included this variant of the classical
approximation (labeled ``CSA'' in Figures \ref{fig:encondaniso2}
and \ref{fig:ratioaniso}) to the comparison with the full calculation,
and it appears that the quadratic terms in $f$ included within this
variant considerably improve the agreement with the full Boltzmann
equation regarding isotropization.

\section{Algorithm}
\label{sec:num}
\subsection{Discretization}
We adopt the following discretization for the longitudinal momentum and the energy:
\begin{itemize}
\item[$\bullet$] The longitudinal momentum $\pz$ is taken in the range
  $[-L,L]$, which we discretize into $2N_z+1$ points (including the
  endpoints $p_z=\pm L$). The longitudinal step $\Delta p_z$ and the
  discrete values $\pz[j]$ (with $j\in[-N_z,N_z]$) are given by
\begin{align}
\Delta p_z=\null&\frac{L}{N_z}& \pz[j]=\null&j\,\Delta p_z\,.
\end{align}

\item[$\bullet$] The energy $\omega_p$ is taken in the range
  $[m+\Delta\omega,\omega_{_\Lambda}]$, which is discretized in $N_{\rm f}$
  points. The step $\Delta\omega$ and the discrete values
  $\omega_p[i]$ (with $i\in[1,N_{\rm f}]$) are given by
\begin{align}
\Delta\omega=\null&\frac{\omega_{_\Lambda}-m}{N_{\rm f}}& \omega_p[i]=\null&i\,\Delta\omega+m\,.
\end{align}

\item[$\bullet$] The transverse momentum $\pp$ is then defined as
\begin{align}
\pp[i,j]=\null&
       \sqrt{\omega^2_p[i]-m^2-\pz^2[j]}& \mbox{if\ \  } \omega^2_p[i]+\pz^2[j]\ge m^2\,,
\label{eq:condonpperp}
\end{align}
with $i\in[1,N_{\rm f}]$ and $j\in[-N_z,N_z]$. If the inequality is
not satisfied, then the pair $(i,j)$ is excluded from the lattice.

\item[$\bullet$] The particle distribution $f(p)$ is encoded as a
  function of $\omega_p$ and $\pz$. In addition, the assumed parity of
  $f$ in $\pz$  translates into
\begin{align}
f[i,j]=\null&f[i,-j]\,.
\end{align}
\end{itemize}
The motivation for this choice is that, with uniformly spaced discrete
energy values, a scattering from a pair of lattice points to a pair
of lattice points will exactly represent energy conservation.  Further,
an integral over all momenta can be represented as a sum over
lattice positions; no sampling or Monte-Carlo integration errors ever
arise, only errors from the discretization procedure itself.

\subsection{Collision term}
\label{sec:Collision term  discretization}
Let us now present an algorithm that preserves all the symmetries of
the collision kernel. This algorithm is a simple extension of the one
used in ref.~\cite{EpelbGTW1} (appendix B.2). For simplicity, we can
assume that $\pza{1}>0$, since the particle distribution is even in
$\pz$. The integrals over $\omega_\p$ and $\pz$ in the collision
kernel are approximated by two 1-dimensional quadrature formulas.  For
the integral of a function $F(\omega_p)$, we use
\begin{align}
\int \d \omega_p\, F(\omega_p) \approx
\Delta\omega\sum_{i=1}^{N_{\rm f}}w_{\rm f}[i]\;F( \omega_p[i])\,.
\end{align}
In our implementation, we have chosen the weights $w_{\rm f}[i]$ as
follows\footnote{Our choice of the quadrature weight $w_{\rm f}[1]$
  assumes that $n_c$ also includes the particles in the energy bin
  $[m,m+\Delta\omega]$.}
\begin{align}
w_{\rm f}[1]=\null&\tfrac{1}{2} &w_{\rm f}[i]=\null&1\quad(i=2\cdots N_{\rm f}-1)&w_{\rm f}[N_{\rm f}]=\null&\tfrac{1}{2}\,.
\end{align}
Similarly, for integrations over the longitudinal momentum $\pz$ we use
\begin{align}
\int \d \pz \,G(\pz)\approx
 \Delta p_z \sum_{j=-N_z}^{N_z} w_z[i]\;G( \pz[i])\,,
\end{align}
with the following weights
\begin{align}
 w_z[-N_z]=\null&\tfrac{1}{2} & w_z[i]=\null&1\quad(i=-N_z+1\cdots N_z-1)& w_z[N_z]=\null&\tfrac{1}{2}\,.
\end{align}

If we denote $(i_1,j_1)$ the integers corresponding to the momentum
$p_1$, the expression (\ref{eq:boltbefdelta}) for the collision kernel
in the absence of condensate can thus be approximated by
\begin{eqnarray}
&&C_{i_1,j_1}[f]
=\frac{g^4\;\left(\Delta\omega\Delta p_z\right)^2}{256\pi^3 \omega_p[i_1]}
\sum_{i_{2,3,4}=1}^{N_{\rm f}} \sum_{j_{2,3,4}=-N_z}^{N_z}
\delta_{i_1+i_2-i_3-i_4}
\delta_{j_1+j_2-j_3-j_4}\nonumber\\
&&\qquad\times\,
w_{\rm f}[i_2]\,w_{\rm f}[i_3]\,w_{\rm f}[i_4]\,
 w_z[j_2]\, w_z[j_3]\, w_z[j_4]
\;
\big[{{\bs I}_4F_{\rm nc}}\big]_{i_{1,2,3,4},j_{1,2,3,4}}\,.
\label{eq:discint}
\end{eqnarray}
In these sums, one should discard any term for which one of the pairs
$(i_a,j_a)$ does not comply with the inequality
(\ref{eq:condonpperp}).  Eqs.~(\ref{eq:cons1}) have been checked to
hold with machine accuracy with this scheme. Similar formulas can be
written for the terms that describe collisions involving a particle
from the condensate, and they are given in appendix \ref{sec:apb}.

\subsection{Free streaming term}
\label{sec:Free-streaming part discretization}
In the absence of collisions (\textsl{e.g.}, in the limit $g^2\to 0$),
the Boltzmann equation describes free streaming, a regime in
which each particle moves on a straight line with a constant
momentum. In the system of coordinates $(\tau,\eta)$, we are
describing a slice in the rapidity variable $\eta$. This slice is
progressively depleted of its particles with a non-zero $\pz$, since
they eventually escape, and the support of the particle distribution
in $\pz$ therefore shrinks linearly with time.

\begin{figure}
\begin{center}
\resizebox*{10cm}{!}{\includegraphics{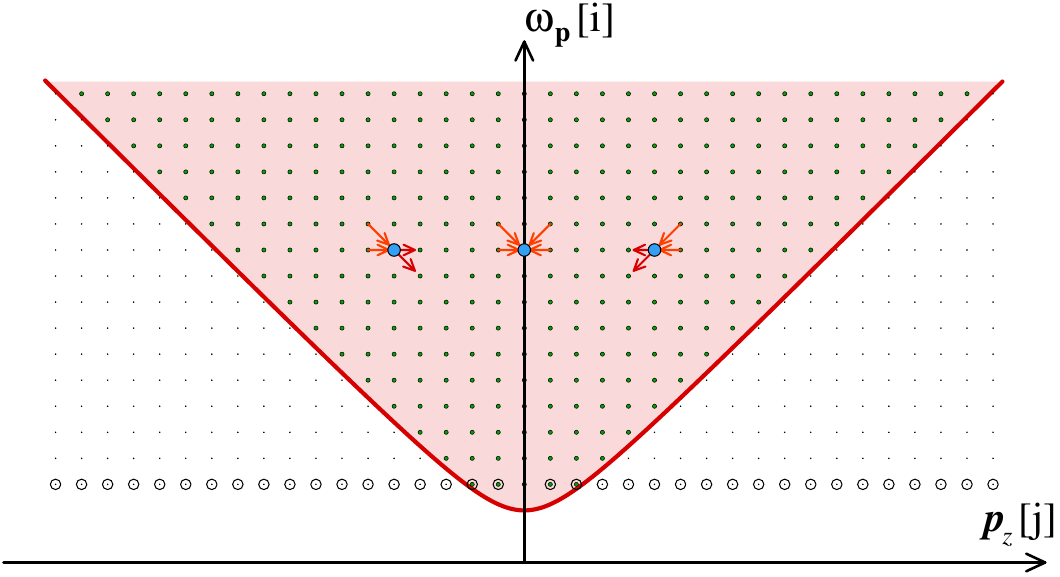}}
\end{center}
\caption{\label{fig:freestrdisc}Possible hops for free streaming
  particles. The shaded area is the kinematically allowed domain (the
  equation of its boundary is $\omega^2=p_z^2+m^2$).}
\end{figure}

On our lattice representing discrete values of $\omega_p$ and $\pz$, the
derivatives with respect to $\omega_p$ and $\pz$ that appear on the
left-hand side of the Boltzmann equation represent the fact that each
particle systematically loses $\pz$, and therefore energy $\omega_p$.
The trajectory of a particle in $(\pz,\omega_p)$ space, shown in
Fig.~\ref{fig:freestrdisc}, will be inward and downward.  Discretizing the
allowed $\pz,\omega_p$ values, this can be viewed as the particles
hopping inward and downward. For instance, the particle number
on the site $(i,j>N_z)$ will move towards the sites%
\footnote{This choice is not unique. For instance, one could instead
 consider hops to $(i,j-1)$ or $(i-1,j)$, which would correspond to a
 different discretization of the derivatives $\partial_\omega$ and
 $\partial_{p_z}$.  We choose this discretization because the $\pz$ change
 is always larger than the $\omega_p$ change, and because the point
 $(i-1,j-1)$ almost always exists, while along the kinematic boundary
 the point $(i-1,j)$ generally does not.}
$(i,j-1)$ and $(i-1,j-1)$. Meanwhile, particle number living on the site
$(i,j+1)$ or $(i+1,j+1)$ will flow onto the site $(i,j)$. In contrast, a
particle on the site $(i,j<N_z)$ (\ie\ $\pz<0$) can hop to $(i,j+1)$
or $(i-1,j+1)$, while a particle located at $(i,j-1)$ or $(i+1,j-1)$
can jump to $(i,j)$. Once a particle reaches the line $j=0$, it does
not move from there. These moves are illustrated in Figure
\ref{fig:freestrdisc}.
Since $f[i,-j]=f[i,j]$, it is sufficient to consider $j\geq 0$. Let us
denote the number density of particles at site $(i,j)$ as
\begin{align}
h[i,j]\equiv \omega_p[i]f[i,j]\,.\label{eq:localn}
\end{align}
The most general form for a discrete version of the collisionless
Boltzmann equation can be written as\footnote{The special
  case $j=0$ is written explicitly in the eq.~(\ref{eq:n10}) of the
  appendix \ref{sec:apc}.}
\begin{eqnarray}
&&\tau\partial_\tau \big(w_{\rm f}[i] w_z[j]\, h[i,j]\big)=
 -\alpha_{ij}\,w_{\rm f}[i]\, w_z[j]\, h[i,j]\nonumber\\
&&\qquad\qquad\qquad\qquad
 +\beta_{ij+1}\,w_{\rm f}[i]\, w_z[j+1]\,h[i,j+1]\nonumber\\
&&\qquad\qquad\qquad\qquad
+\gamma_{i+1j+1}\,w_{\rm f}[i+1]\, w_z[j+1]\, h[i+1,j+1] \label{eq:n1}\,,
\end{eqnarray}
where $(\alpha,\beta,\gamma)_{ij}$ are coefficients that will
be adjusted in order to satisfy all the conservation laws\footnote{We
  can disregard the condensate in this subsection. Indeed, since the
  particles in the condensate have zero momentum, they play no role in
  free streaming, which simply causes the condensate number density
  to decay as $\tau^{-1}$.}.

The non-condensed contribution to the particle density reads
\begin{align}
n_{\rm nc}=\sum_{i=1}^{N_{\rm f}}\sum_{j=-N_z}^{N_z}w_{\rm f}[i]\, w_z[j]\;h[i,j]\,.\label{eq:discpartnum}
\end{align}
Similarly, we can write its contribution to the energy density and
longitudinal pressure as follows
\begin{eqnarray}
\epsilon_{\rm nc}&\smeq&\sum_{i=1}^{N_{\rm f}}\sum_{j=-N_z}^{N_z}w_{\rm f}[i]\, w_z[j]\,\omega_p[i]\,h[i,j]\,,\label{eq:discener}
\\
P_{_L{\rm nc}}&\smeq&\sum_{i=1}^{N_{\rm f}}\sum_{j=-N_z}^{N_z}w_{\rm f}[i]\, w_z[j]\,\frac{\pz^2[j]}{\omega_p[i]}\,h[i,j]\, .
\label{eq:discPL}
\end{eqnarray}
In order to fully determine the unknown coefficients, we also need to
consider the first moment of the distribution of longitudinal momenta,
\begin{align}
\rho_{z}\equiv \sum_{i=1}^{N_{\rm f}}\sum_{j=-N_z}^{N_z}w_{\rm f}[i]\, w_z[j]\,\pz[j]\,h[i,j]\,.\label{eq:discfirstmoment}
\end{align}
In the appendix \ref{sec:apc}, we show that
$(\alpha,\beta,\gamma)_{ij}$ must have the following form\footnote{A
  limitation of our scheme is that it works only if the points
  $(i=1,j=\pm 1,\pm2,\cdots)$ are not allowed by the condition
  $\omega_p^2>m^2+\pz^2$, \ie\ if
\begin{equation*}
\big(\Delta\omega\big)^2+2m\Delta\omega<\big(\Delta p_z\big)^2\,.
\end{equation*}
This can be fulfilled by choosing appropriately the lattice parameters
(these points have been surrounded by a circle in Figure
\ref{fig:freestrdisc} -- in the example of this figure, the above
inequality is not satisfied). All the numerical results shown in this
paper have been obtained with a lattice setup that satisfies this
condition.}:
\begin{align}
\alpha_{ij}=\null&1+\frac{\pz[j]}{\Delta p_z}\,,&
\beta_{ij}=\null&2\frac{\pz[j+1]}{\Delta p_z}-\frac{\pz^2[j]}{\omega_p[i]\Delta\omega}\,,&
\gamma_{ij}=\null&\frac{\pz^2[j]}{\omega_p[i]\Delta\omega}\,.
\label{eq:coeffsol}
\end{align}
Starting from (\ref{eq:n1}) and replacing the local particle density
by (\ref{eq:localn}), we obtain the following discretization (for $j>0$)
for the free streaming equation
\begin{eqnarray}
\partial_\tau f[i,j]&\smeq&
-\frac{1}{\tau}\left(1+\frac{\pz[j]}{\Delta p_z}\right)f[i,j]\nonumber\\
&&
+\frac{1}{\tau}\frac{ w_z[j+1]}{  w_z[j]}\left(\frac{\pz[j+1]}{\Delta p_z}-\frac{\pz^2[j+1]}{\omega_p[i]\Delta\omega}\right)
 f[i,j+1]\nonumber\\
&&+\frac{1}{\tau}\frac{w_{\rm f}[i+1]}{w_{\rm f}[i]}
\frac{ w_z[j+1]}{ w_z[j]}\frac{\pz^2[j+1]}{\omega_p[i]\Delta\omega}
f[i+1,j+1]\,.\label{eq:fkdiscfs}
\end{eqnarray}
In the particular  case $j=0$, this equation reads
\begin{align}
\partial_\tau f[i,0]\null&=-\frac{1}{\tau}f[i,0]
+\frac{2}{\tau}\frac{ w_z[1]}{  w_z[0]}\left(1-\frac{(\Delta p_z)^2}{\omega_p[i]\Delta\omega}\right)
 f[i,1]\notag\\
&+\frac{2}{\tau}\frac{w_{\rm f}[i+1]}{w_{\rm f}[i]}
\frac{ w_z[1]}{ w_z[0]}\frac{(\Delta p_z)^2}{\omega_p[i]\Delta\omega}
f[i+1,1]\,.
\end{align}
One can also rewrite eq.~(\ref{eq:fkdiscfs}) as
\begin{eqnarray}
&&
\partial_\tau f[i,j]=\frac{\pz[j+1]}{\tau}\left(\frac{ w_z[j+1] f[i,j+1]- w_z[j]f[i,j]}{\Delta p_z  w_z[j]}\right)\nonumber\\
&&\qquad
+\frac{ w_z[j+1]}{  w_z[j]}\frac{\pz^2[j+1]}{\tau\omega_p[i]}
 \left(\frac{w_{\rm f}[i+1]f[i+1,j+1]-w_{\rm f}[i]f[i,j+1]}{\Delta\omega}\right)
\,.\nonumber\\
&&\label{eq:fkdiscfs23}
\end{eqnarray}
For the internal points, where all the weights $w_{\rm f}[i]$ and
${\rm w}_{z}[j]$ are equal to one, this becomes
\begin{eqnarray}
\partial_\tau f[i,j]&\smeq&
\frac{\pz[j+1]}{\tau}\left(\frac{ f[i,j+1]-f[i,j]}{\Delta p_z}\right)
\nonumber\\
&&
+\frac{\pz^2[j+1]}{\tau\,\omega_p[i]}
 \left(\frac{f[i+1,j+1]-f[i,j+1]}{\Delta\omega}\right)
\,,
\end{eqnarray}
which indeed reproduces the left-hand side of the Boltzmann equation
(\ref{eq:Bfc}) in the continuum limit.

\section{Numerical results}
\label{sec:Resolution of the Boltzmann equation}
\subsection{CGC-like initial condition}
We now solve the coupled equations (\ref{eq:Bfc}) and (\ref{eq:Cc1})
with the algorithm described in the previous section.  We have used a
moderately anisotropic initial condition that mimics the gluon
distribution in the Color Glass Condensate at a proper time $\tau\sim
Q_s^{-1}$. It is characterized by a single momentum scale $Q$, below
which most of the particles lie. The scale $Q$ also sets the unit for
all the other dimensionful quantities. To be more specific, our
initial distribution at the initial time $Q\tau_0=1$ is:
\begin{align}
  f_{{\rm init}}(\omega_p,\pz)=\null f_0\;\exp\left(-\alpha\,
    \tfrac{\omega_p^2}{Q^2}-\beta\, \tfrac{\pz^2}{Q^2}\right)\,,
\label{eq:finitdics}
\end{align}
with a large occupation below $Q$, $f_0=100$. According to the
standard argument, such a large value of $f_0$ should ensure that the
classicality condition is well satisfied, and that the cubic terms
alone lead to good approximation of the full solution. We choose a
coupling constant\footnote{\label{foot:coupling}Although this may seem
  to be a large value, it corresponds to a rather small scattering
  rate, because of the prefactor $g^4/(256\pi^3)$ in front of the
  collision integral.  Another point of view on this value is to
  recall that the screening mass in a $\phi^4$ scalar theory at
  temperature $T$ is $m_{\rm scr}^2=g^2T^2/24$, while in Yang-Mills
  theory with 3 colors it is $m_{\rm scr, YM}^2=g^2_{_{\rm
      YM}}T^2$. Thus, if the two theories were compared at equal
  screening masses, one would have $g^2=24\,g^2_{_{\rm
      YM}}$. Alternatively, if we compare the two theories at the same
  shear viscosity \cite{Jeon1,Jeon2,ArnolMY6} to entropy density
  ratio, the scalar and gauge couplings should be related by
  $g^2\approx 40\, g_{_{\rm YM}}^2\big({\log(g_{_{\rm
        YM}}^{-1})}\big)^{1/2}$. A coupling $g^4=50$ in the scalar
  theory would correspond to a very small strong coupling constant
  $\alpha_s\sim 0.023$ (conversely, $\alpha_s=0.3$ would correspond to
  choosing $g^4\sim 10^4$ in the scalar theory). Note that if $g^4=50$
  is the coupling at the scale $Q$, then the Landau pole of the
  $\phi^4$ theory is at the scale $\mu=Q\,\exp(16\pi^3/(3 g^2))\approx
  1844\,Q$ -- sufficiently above $Q$ to justify a perturbative
  treatment.}  $g^4=50$. The particle density in the condensate is
initially $n_{c,{\rm init}}=10^{-6}$, and the mass $m$ is taken to be
$m/Q=0.1$. Finally, the cutoffs are $L/Q=5$ and
$\omega_{_\Lambda}=\sqrt{L^2+m^2}$, while $N_{\rm f}=2N_z=64$.  The
initial anisotropy was moderate, controlled by the parameters
$\alpha=2$ and $\beta=4$.

\begin{figure}[htbp]
\begin{center}
\resizebox*{9cm}{!}{\includegraphics{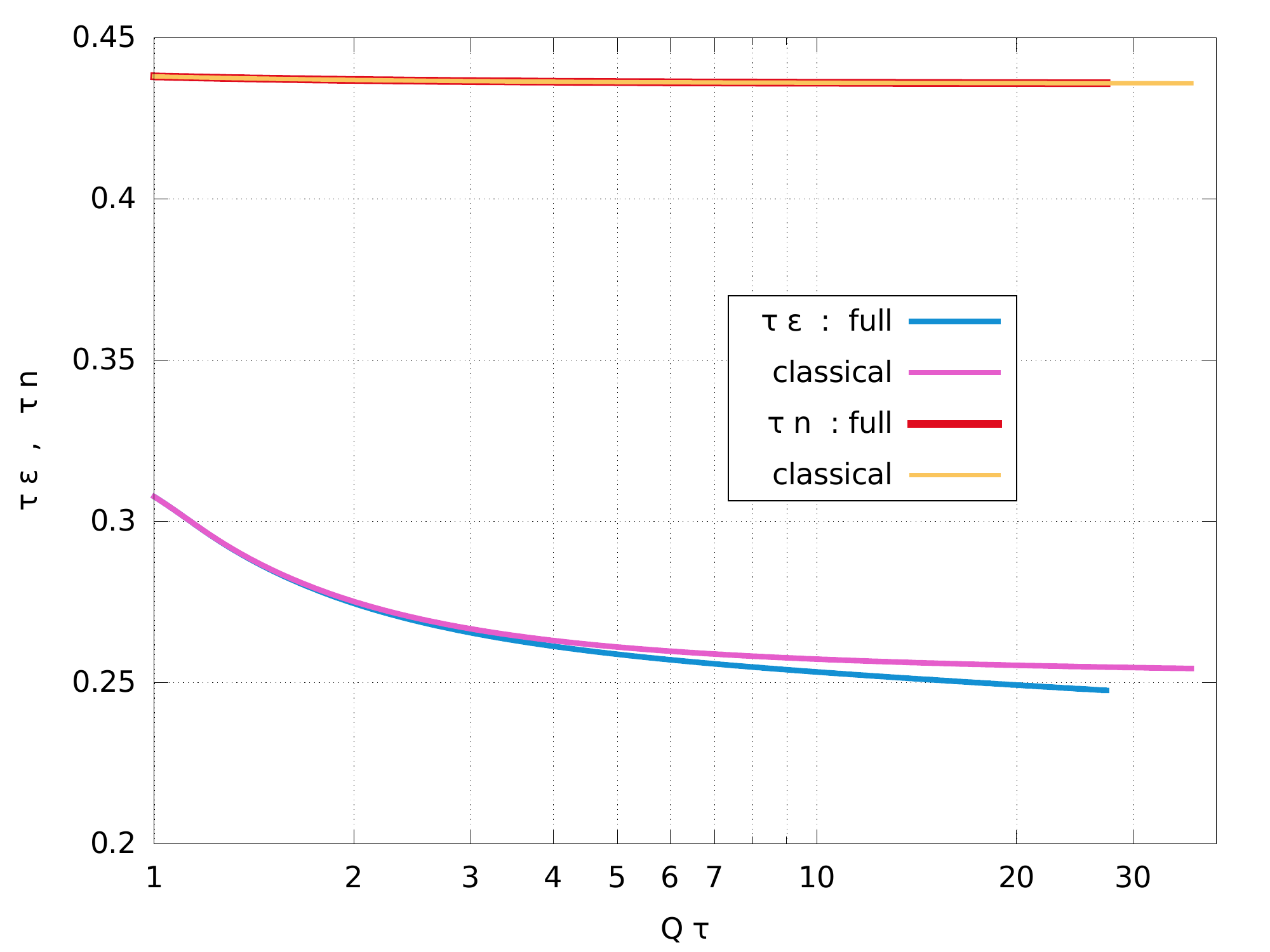}}
\end{center}
\caption{\label{fig:rencondexpand}Proper-time evolution of the
  energy-density and particle number times $\tau$ as defined in
  (\ref{eq:discpartnum}) and (\ref{eq:discener}) for the different
  schemes.}
\end{figure}

In Figure \ref{fig:rencondexpand}, we show the time evolution of
$\tau n$ and $\tau \epsilon$ in the unapproximated and in the
classical schemes. $\tau n$ should be strictly constant\footnote{Our
  discretization of the momentum integrals ensures exact conservation
  equations only if the time derivatives are evaluated exactly. The
  numerical resolution of the Boltzmann equation therefore also
  introduces an error that depends on the timestep $\Delta\tau$ and on
  the details of the scheme used for the time evolution. With our
  implementation, the quantity $(\tau-\Delta\tau)n(\tau)$ is conserved
  with machine precision, and the expected conservation law is exactly
  recovered in the limit $\Delta\tau\to 0$.} in both cases, since the
conservation of particle number is not affected by the classical
approximation (thus, this quantity is just used to monitor how well
this conservation law is satisfied in the numerical implementation).
A small difference between the two schemes is visible in the energy
density. Given the conservation equation (\ref{eq:T-cons}), this also
indicates that the two schemes lead to different longitudinal
pressures. Since the unapproximated scheme leads to a faster decrease
of the energy density than the classical scheme, it must have a larger
longitudinal pressure.  This will be discussed in greater detail later
in this section.

\subsection{Bose-Einstein condensation}
\begin{figure}[htbp]
\begin{center}
\resizebox*{9cm}{!}{\includegraphics{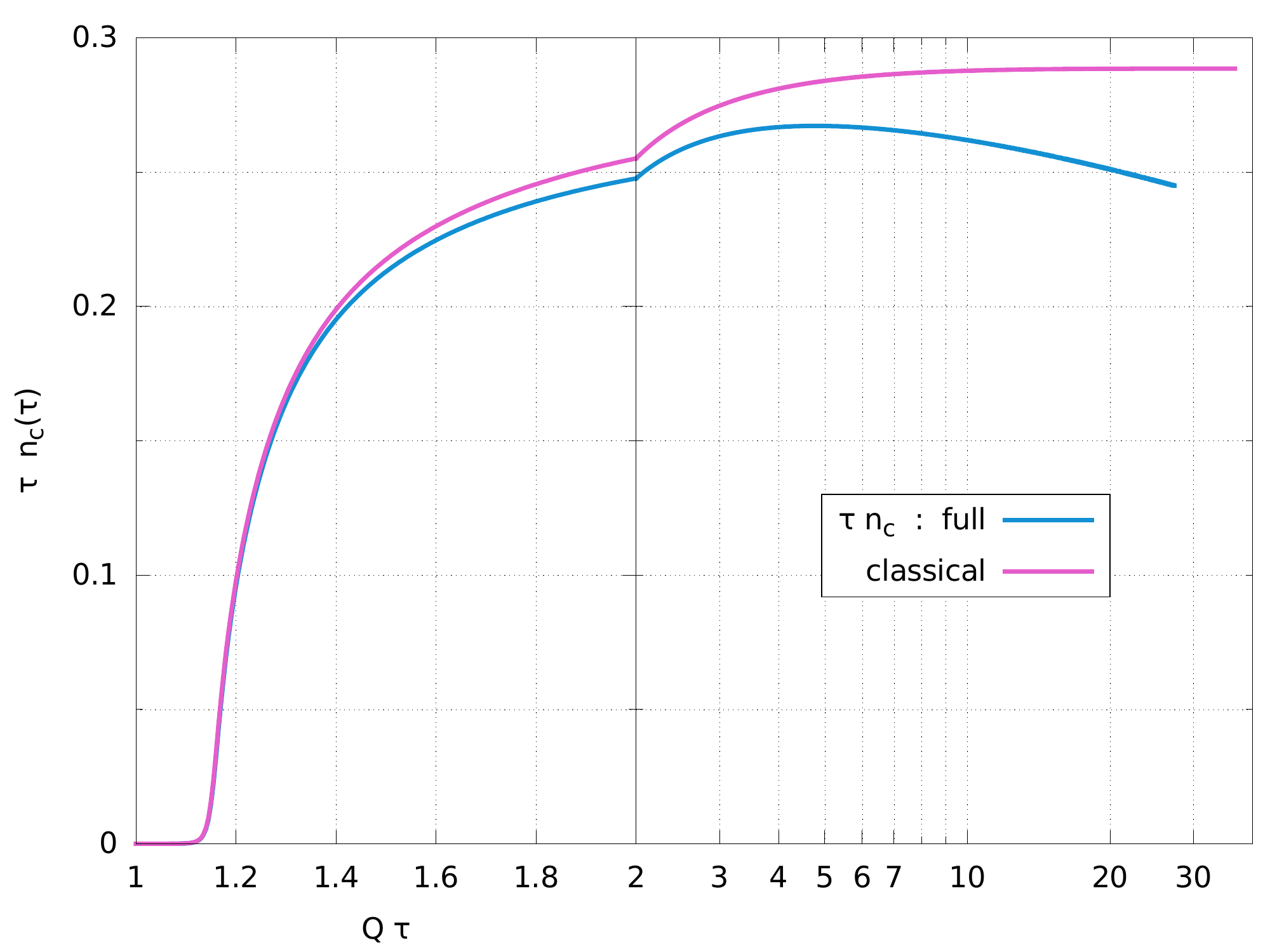}}
\end{center}
\caption{\label{fig:rencondexpand2}Proper-time evolution of the
  condensate times $\tau$ for the different schemes. Note the
  discontinuity in how we plot time at $Q\tau=2$ (linear scale on the
  left and logarithmic scale on the right), which artificially causes
  a cusp in the curves.}
\end{figure}

The initial condition that we have chosen corresponds to a large
overpopulation, since $[n\epsilon^{-3/4}]_{\tau_0}\gg 1$. If the
system were not expanding, we would expect the formation of a
Bose-Einstein condensate. Figure \ref{fig:rencondexpand2} shows
the particle density in the zero mode in the unapproximated and
classical schemes.
The onset of Bose-Einstein condensation is nearly identical in the two
schemes, and a moderate difference develops at later times, that
reaches about 20\% at $Q\tau\sim 10$. The rather small
difference between the two schemes for this quantity can be understood
from the fact that the evolution of the condensate is governed by the
region of small momenta, where the particle distribution is very
large.  We also see here a trend already observed in the isotropic
case in ref.~\cite{EpelbGTW1}: the classical approximation leads to
more condensation than the unapproximated collision term. In fact,
when the ultraviolet cutoff is large compared to the physical momentum
scales, most of the particles tend to aggregate in a condensate in the
classical approximation.

We mention in passing that our algorithm is not particularly well
suited for a detailed study of the infrared region.  The fixed energy
spacing of our discretization lacks resolution in the IR, and our
treatment of the mass as fixed, rather than a self-consistently
determined thermal mass, is another limitation.  On the other hand,
the infrared has the highest occupancies, so classical methods are
most reliable there.  Therefore, lattice classical field simulations
are much better suited for studying the infrared region.  In
particular our method is too crude to reveal the interesting scaling
regimes found for instance in Ref.~\cite{BergeBSV3}.  For this reason
we will concentrate on quantities which are controlled by the
higher-energy excitations, such as the components of the pressure.

\subsection{Pressure anisotropy}
\begin{figure}[htbp]
\begin{center}
\resizebox*{9cm}{!}{\includegraphics{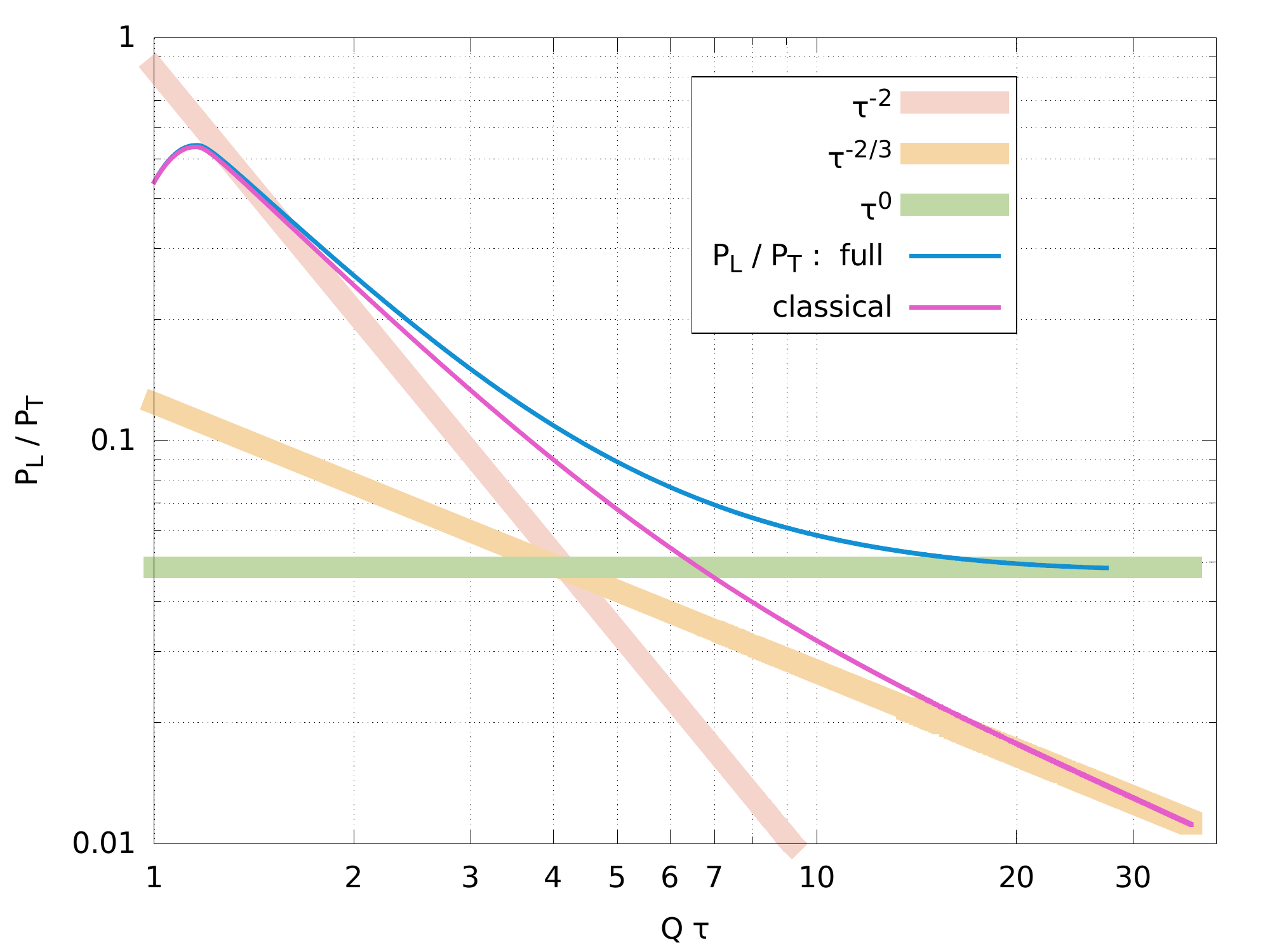}}
\end{center}
\caption{\label{fig:ratio_expand}Time evolution of ${P_{_L}}/{P_{_T}}$.}
\end{figure}

\begin{figure}[htbp]
\begin{center}
\resizebox*{9cm}{!}{\includegraphics{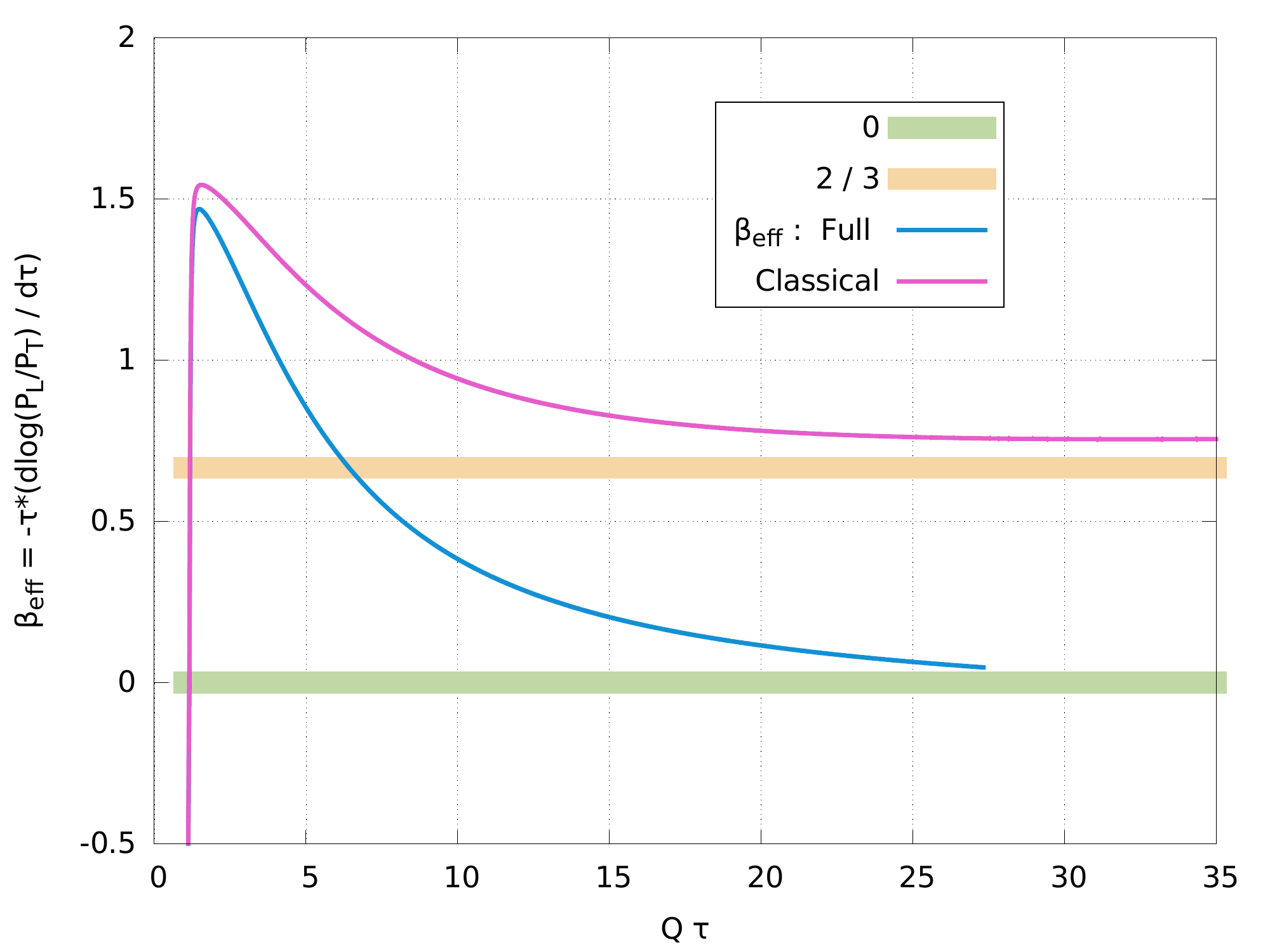}}
\end{center}
\caption{\label{fig:tau_alpha}Time evolution of $\beta_{\rm eff}\equiv-\tau\, \d\ln({P_{_L}}/{P_{_T}})/\d\tau$.}
\end{figure}

More important differences between the two schemes can be seen in the
behavior of the longitudinal pressure. In Figure
\ref{fig:ratio_expand}, we display the time evolution of the ratio
${P_{_L}}/{P_{_T}}$. The beginning of the evolution is similar in the
two schemes, with a brief initial increase of this ratio due to
scatterings. Rapidly, the expansion of the system takes over and makes
the ratio decrease, but at a pace slower than free streaming
(indicated by a band falling like $\tau^{-2}$). The two schemes start
behaving differently around $Q\tau\sim 2$, with the classical
approximation leading to a faster decrease of the ratio
${P_{_L}}/{P_{_T}}$, approximately like $\tau^{-2/3}$. In contrast,
the unapproximated collision term seems to lead to a constant ratio at
large times.
In Figure \ref{fig:tau_alpha}, we display the quantity $\beta_{\rm
  eff}\equiv-\tau\, \d\ln({P_{_L}}/{P_{_T}})/\d\tau$ as a function of
time. If we parameterize
\begin{equation}
\frac{P_{_L}}{P_{_T}}=C\cdot\big(Q\tau\big)^{-\beta(\tau)}\, ,
\end{equation}
and if the exponent $\beta$ is slowly varying, then $\beta_{\rm eff}$
gives the instantaneous value of this exponent. This figure is to be
compared with Figure \ref{fig:iso}, where several scenarios for
the behavior of this exponent have been presented.
On this plot, we see that this exponent behaves quite differently in
the two schemes, the asymptotic exponent being close to $2/3$ in the
classical approximation, while it is zero with the unapproximated
collision term. Moreover, the exponent $2/3$ does not appear to play
any particular role when one uses the full collision term, since
$\beta_{\rm eff}$ does not spend any time at this value in this case,
despite the large occupation number in this
simulation. Therefore, the classical attractor scenario represented by
the red curve in Figure \ref{fig:iso} is not realized for this
combination of initial condition and coupling. This computation also
indicates that the condition $f\gg 1$, that was regarded as a
criterion for classicality, should be used with caution. In this
example, it does not guarantee that the classical approximation
describes correctly the evolution of the system. This condition is
imprecise because $f$ is in fact a function of momentum, and $f\gg 1$
may not be true over all the regions of phase-space that dominate the
collision integral.

Note that if the ratio $P_{_L}/\epsilon$ is approximately
constant,
\begin{equation}
P_{_L}=\delta\,\epsilon\, ,
\end{equation}
(as is the case in the full calculation at large times) then we have
\begin{equation}
n\sim \tau^{-1}\quad,\qquad\epsilon \sim \tau^{-(1+\delta)}\, ,
\end{equation}
and the overpopulation measure behaves as follows:
\begin{equation}
n\epsilon^{-3/4}\sim \tau^{\tfrac{3\delta-1}{4}}\, .
\end{equation}
For an isotropic system, $\delta=1/3$ and $n\epsilon^{-3/4}$ is a
constant, and the Bose-Einstein condensate would survive forever (in
our kinetic approximation where inelastic processes are not
included). If the system remains anisotropic at large times, we have
$\delta<1/3$ and $n\epsilon^{-3/4}$ decreases. Therefore, one expects
that, if a condensate forms, it has a finite lifetime because the
overpopulation condition will not be satisfied beyond a certain
time. The final outcome should therefore be the disappearance of the
condensate. The beginning of this process is visible in Figure
\ref{fig:rencondexpand2} in the case of the unapproximated collision
term.

\begin{figure}[htbp]
\begin{center}
\resizebox*{9cm}{!}{\includegraphics{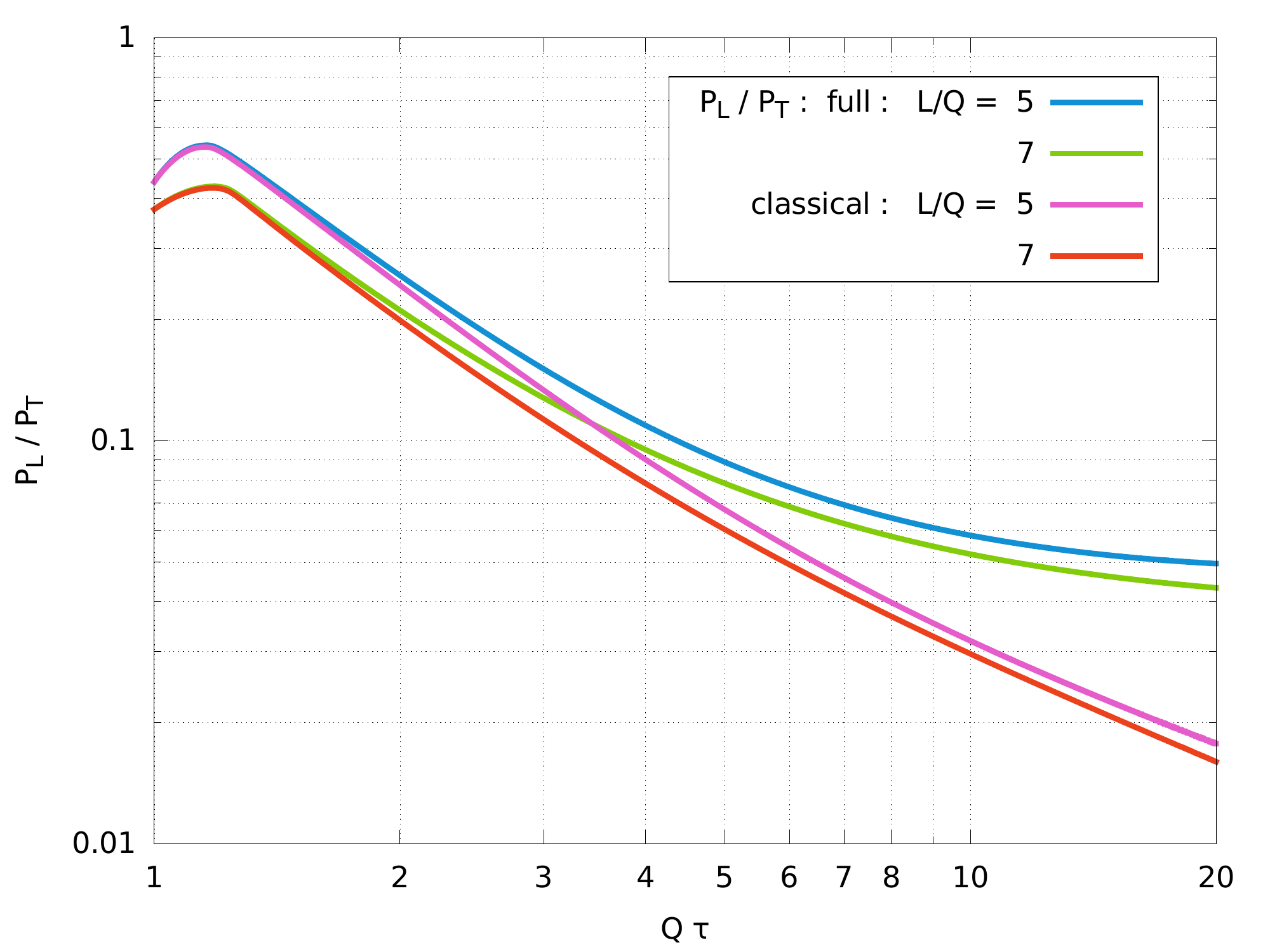}}
\end{center}
\caption{\label{fig:pzdep}Time evolution of ${P_{_L}}/{P_{_T}}$ for
  two values of the cutoff on $\pz$.}
\end{figure}

We have also investigated the sensitivity of our algorithm to the
ultraviolet cutoff $L$ on the longitudinal momentum. This cutoff may
indeed have an important influence on the result since the typical
$\pz$ of the particles in the system evolves with time due to the
expansion. In Figure \ref{fig:pzdep}, we compare the results for
$L/Q=5$ and $L/Q=7$. We observe that the difference between the two
cutoffs is essentially the one inherited from the initial condition,
\ie\ the fact that the tail of the Gaussian in
eq.~(\ref{eq:finitdics}) extends further when we increase the cutoff.
The qualitative differences between the classical and full results are
independent of the value chosen for this cutoff.

\begin{figure}[htbp]
\begin{center}
\resizebox*{9cm}{!}{\includegraphics{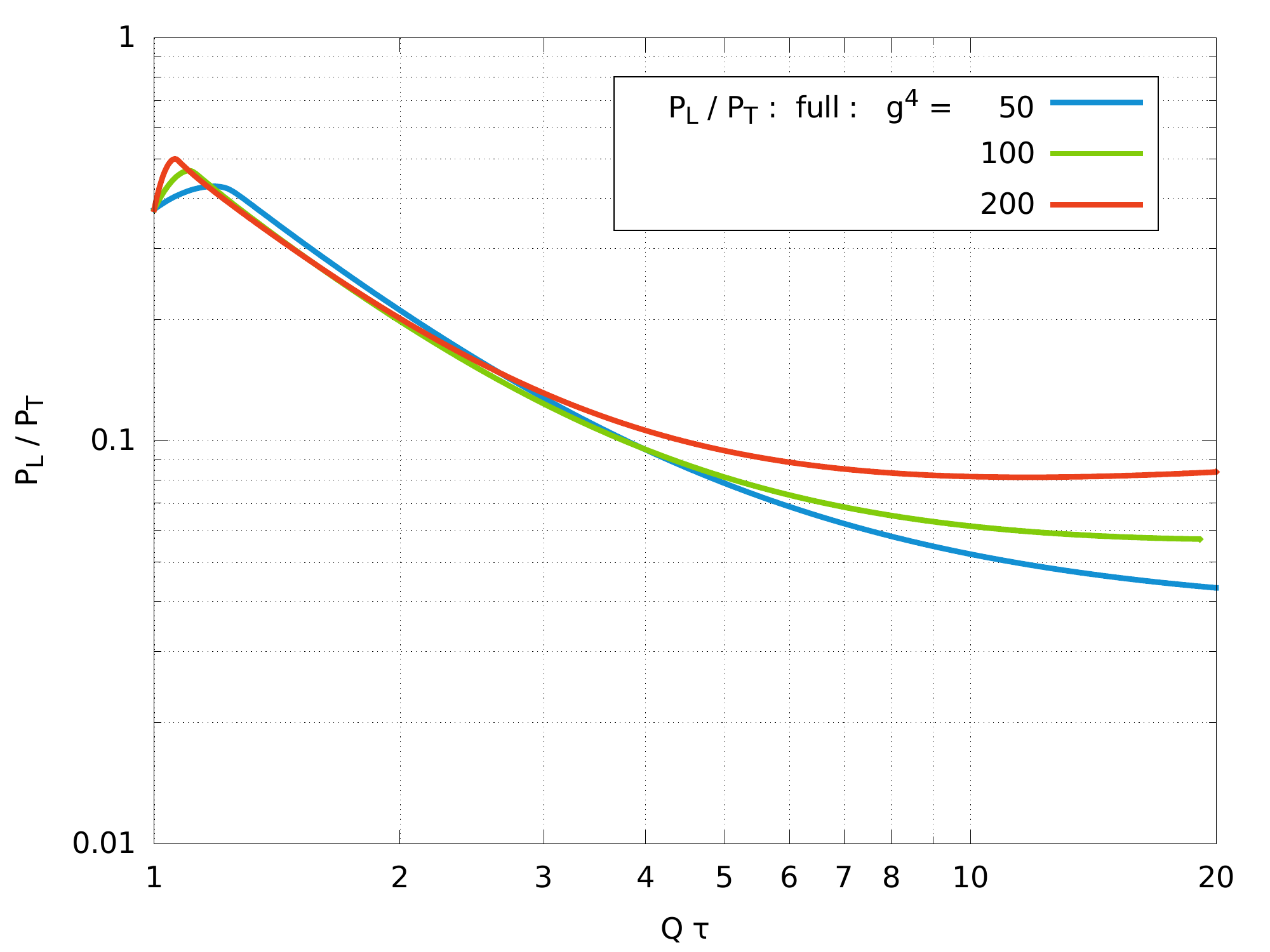}}
\end{center}
\caption{\label{fig:gdep}Time evolution of ${P_{_L}}/{P_{_T}}$ for
  several values of the coupling.}
\end{figure}

In Figure \ref{fig:gdep}, we vary the coupling constant in order to
see how the asymptotic behavior of the full solution is affected by
the strength of the interactions.
For the three values of the coupling, the ratio ${P_{_L}}/{P_{_T}}$
reaches a minimum, whose value increases with the coupling. For the
largest of the couplings we have considered ($g^4=200$, \ie\ $g\approx
3.76$), this ratio even shows a slight tendency to increase after a
time of order $Q\tau\approx 12$.

\subsection{More results using the DSMC algorithm}
The deterministic algorithm we have used so far provides a direct
resolution of the Boltzmann equation, but requires at each timestep
the very time consuming computation of the collision
integral. Moreover, it has a rather unfavorable scaling with the
number of lattice points used in order to discretize momentum space.
For this reason, we have also implemented a stochastic algorithm, the
``direct simulation Monte-Carlo'' (DSMC, described in the appendix
\ref{app:dsmc}), where the distribution $f$ is replaced by a large
ensemble of simulated particles. By construction, energy, momentum and
particle number are exactly conserved with this algorithm (provided
the kinematics of the collisions is treated exactly). Its sources of
errors are the limited statistics, and the reconstruction of the
particle distribution from the simulated particles (this step requires
a discretization of momentum space, which leads to some additional
errors).

\begin{figure}[htbp]
\begin{center}
\resizebox*{9cm}{!}{\includegraphics{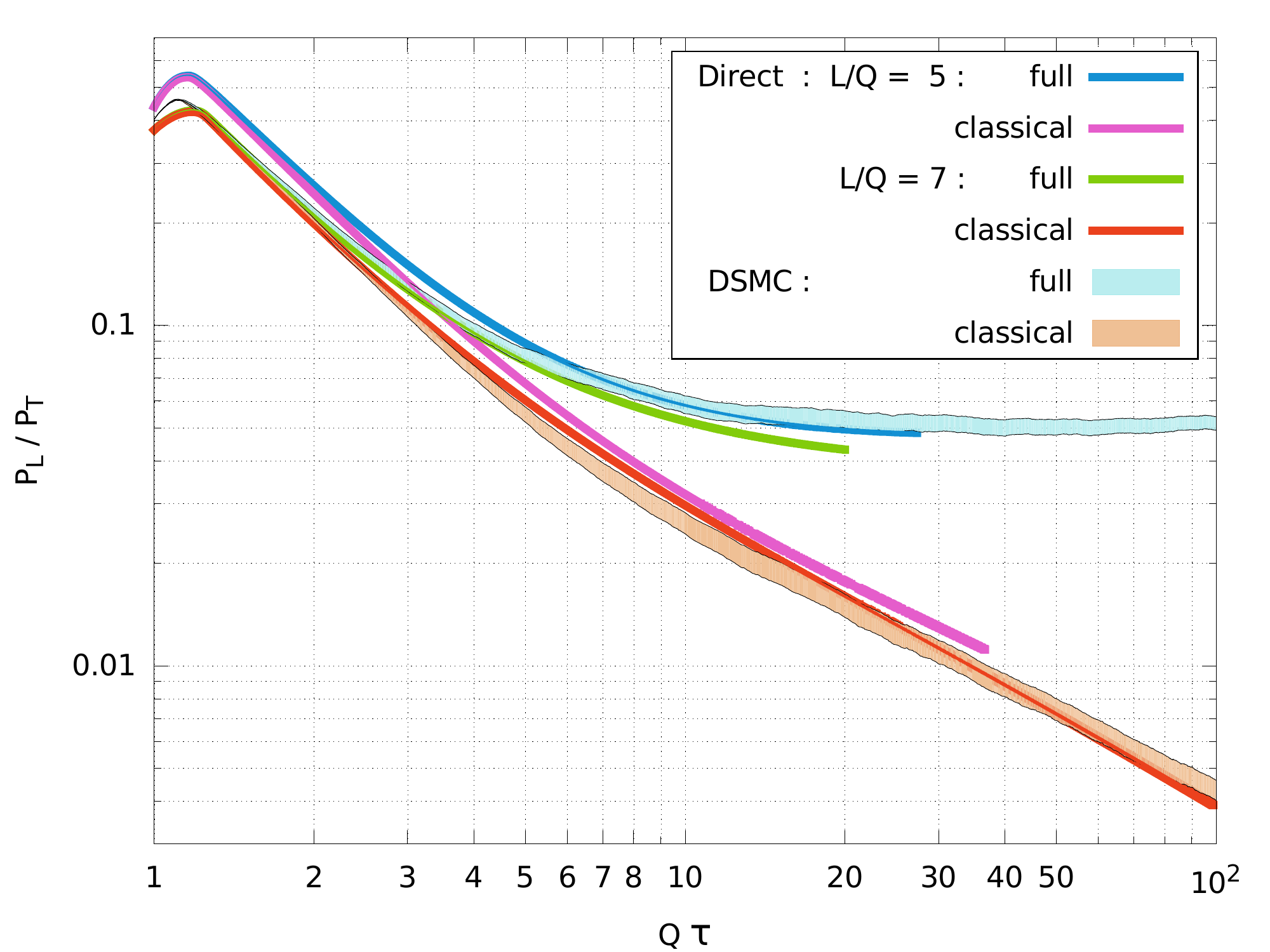}}
\end{center}
\caption{\label{fig:dsmc}Comparison of the deterministic method
  (``Direct'') and the DSMC algorithm (see the appendix
  \ref{app:dsmc}), for the initial condition (\ref{eq:finitdics}) with
  $\alpha=2$, $\beta=4$ and $g^4=50$. {In the DSMC case, the band is
    an estimate of the systematic error based on the different values
    of the pressure one obtains by including or not the particles from
    the ``condensate'' (since the definition of the condensate in the
    DSMC includes all particles in a small volume around $\p=0$).}}
\end{figure}

Before showing more results using this algorithm, we have first used
it with the same initial condition already used with the deterministic
method, in order to compare the two approaches. The outcome of this
comparison is shown in Figure \ref{fig:dsmc}, and indicates a good
agreement between the two methods.
The differences, in the 10~\% to 20~\% range, can be attributed to the
fact that the deterministic method simply disregards the particles
with momenta higher than the lattice cutoffs. In contrast, in the DSMC
method, there is no limit on the momenta of the simulated particles,
and a discretization of momentum space is only used when
reconstructing the distribution $f$ from the ensemble of particles.

We have then used the DSMC algorithm in order to study the time
evolution of the pressure ratio $P_{_L}/P_{_T}$ in two situations:


First, we fix the value of the coupling constant at $g^4=50$,
  and we vary the prefactor $f_0$ in the initial condition given by
  eq.~(\ref{eq:finitdics}). The parameters $\alpha$ and $\beta$ that
  control the initial anisotropy of this distribution are also held
  fixed, as well as the initial time $Q\tau_0=1$.
\begin{figure}[htbp]
\begin{center}
\resizebox*{9cm}{!}{\includegraphics{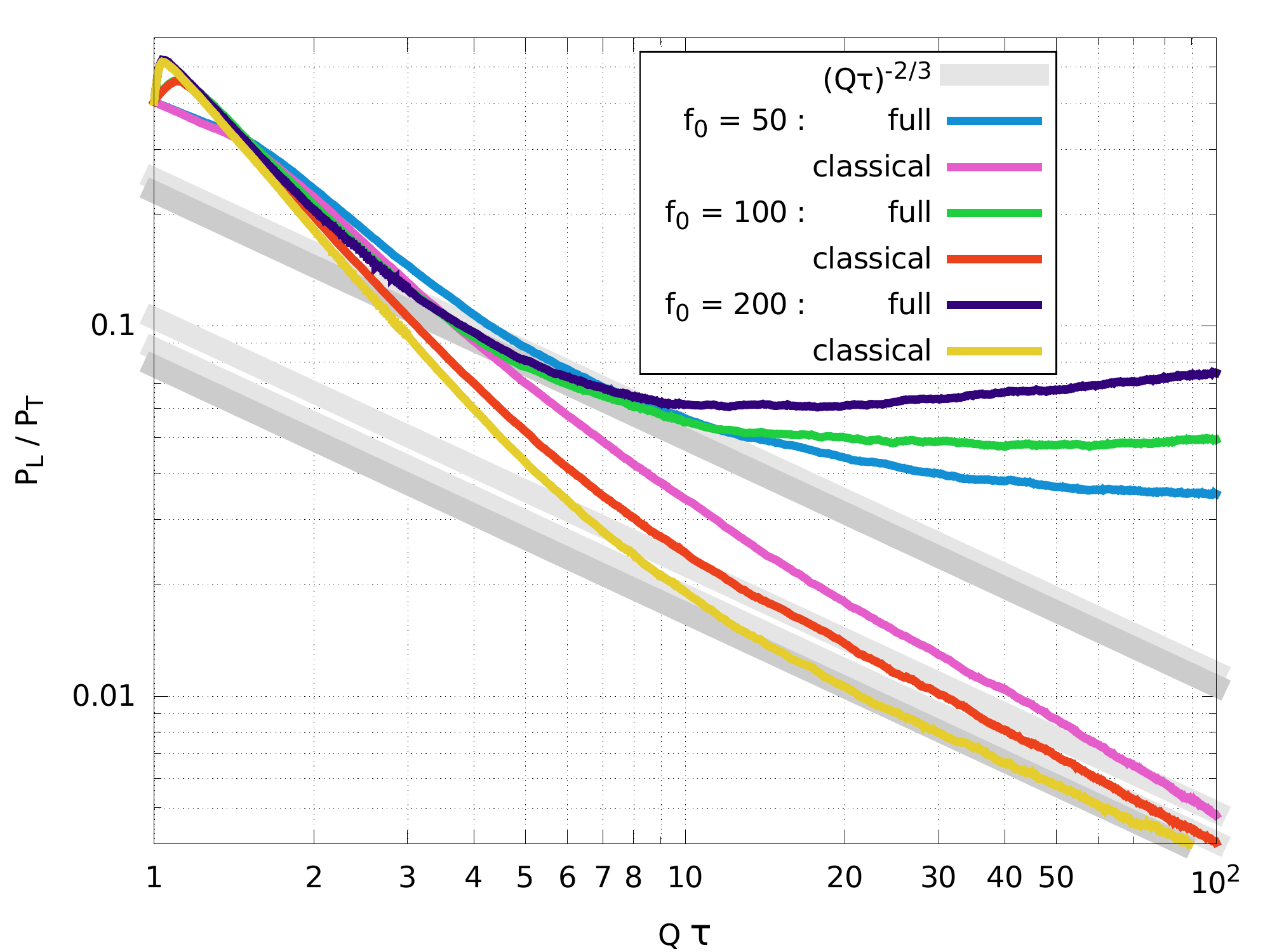}}
\end{center}
\caption{\label{fig:f0}Ratio $P_{_L}/P_{_T}$ as a function of time,
  for a fixed coupling $g^4=50$ and various amplitudes of the initial
  occupation number $f_0=50,100,200$, with the full collision term and
  in the classical approximation. The gray bands indicate the
  classical attractor behavior $(Q\tau)^{-2/3}$.}
\end{figure}
Although one may naively expect the agreement between the full and
classical results to improve when $f_0$ is increased, Figure
\ref{fig:f0} shows that this is not the case. For the three values of
$f_0$ considered in this computation, the classical approximation
departs from the full result at roughly the same early time (or even a
little earlier for the largest $f_0$). No matter how large $f_0$, the
ratio $P_{_L}/P_{_T}$ becomes roughly constant at late times --or even
slightly increases-- in the full calculation, and decreases like
$(Q\tau)^{-2/3}$ in the classical approximation.

Next, in a second series of computations, we have varied
  simultaneously the coupling constant and the initial occupation
  number in such a way that $g^2 f_0$ remains constant, $g^2
  f_0\approx 700$. The parameters $\alpha$ and $\beta$ controlling the
  initial anisotropy are the same as before.
\begin{figure}[htbp]
\begin{center}
\resizebox*{9cm}{!}{\includegraphics{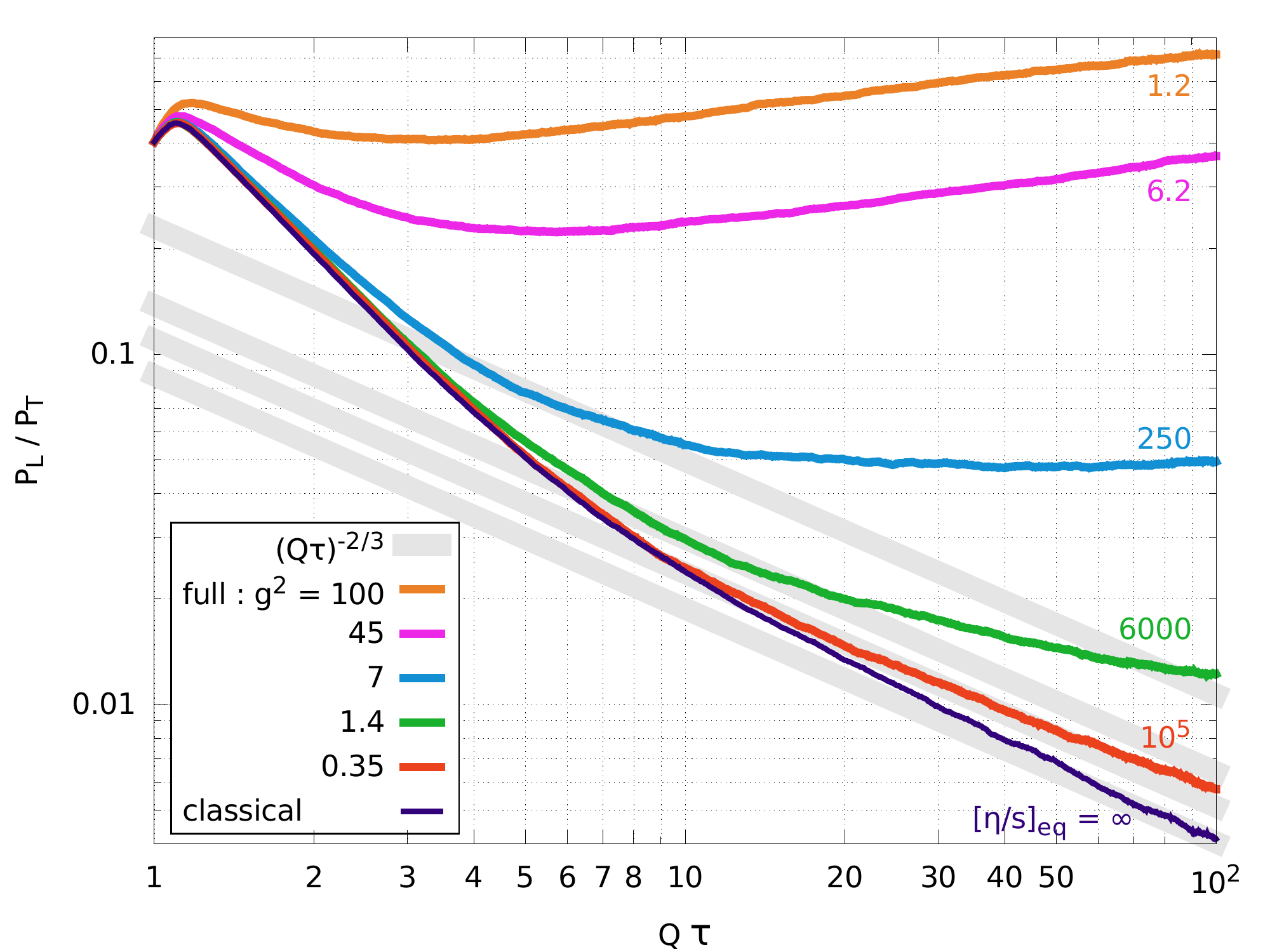}}
\end{center}
\caption{\label{fig:g2f0}Ratio $P_{_L}/P_{_T}$ as a function of time,
  for a fixed value of $g^2f_0\approx 700$ and various couplings
  $g^2=0.35, 1.4, 7, 45$ and $100$, with the full collision term and
  in the classical approximation. The gray bands indicate the
  classical attractor behavior $(Q\tau)^{-2/3}$. The numbers overlaid
  on the right indicate the equilibrium value of the ratio $\eta/s$
  (at leading order -- see refs.~\cite{Jeon1,Jeon2}) for the corresponding
  $g^2$.}
\end{figure}
The resulting evolution of the ratio $P_{_L}/P_{_T}$ is shown in the
figure \ref{fig:g2f0}. In the classical approximation, the pressure
ratio falls like $(Q\tau)^{-2/3}$, as already observed earlier. Note
that we have represented only one classical curve, common to all the
values of $g^2$. Indeed, since the collision term in this
approximation is homogeneous in $f$, one can factor out a prefactor
$f_0^2$ and combine it with the $g^4$ of the squared matrix element,
so that the classical dynamics is always the same if $g^2f_0$ is held
fixed. In contrast, the full dynamics does not posses this invariance,
but appears to converge towards the classical result when $g^2\to 0$.
At finite coupling, the agreement between the full and classical
results is good only over a finite time window, that shrinks as $g^2$
increases. At moderate values of the coupling such as $g^2=7$
(corresponding to $g^4=50$, see Footnote \ref{foot:coupling}), the
unapproximated evolution
departs from the classical one at a rather early point in time, and
the exponent $-2/3$ does not play any particular role in the evolution
of $P_{_L}/P_{_T}$. Two larger values of the coupling ($g^2=45$ and
$g^2=100$) are also shown on this plot, but one should not take the
corresponding results seriously.  Indeed, scalar theory with such a
large coupling is not really self-consistent, because the coupling
runs very fast and the Landau pole is only a factor of 5 to 20 away.


\section{Summary and conclusions}
\label{sec:concl}
This paper started with the qualitative observation that large-angle
out-of-plane scatterings are artificially suppressed by the classical
approximation of the collision term in the Boltzmann equation with
$2\to 2$ scatterings, when the particle distribution is anisotropic,
as is generically the case for a system subject to a fast longitudinal
expansion. This kinematics is for instance realized in the early
stages of heavy ion collisions.

In order to quantify this effect, we have considered a longitudinally
expanding system of real scalar fields with a $\phi^4$ interaction in
kinetic theory, and we have solved numerically the Boltzmann equation
with elastic scatterings in two situations: (a) with the full
collision term, and (b) in the classical approximation where one keeps
only the terms that are cubic in the particle distribution.

This numerical resolution has been performed with two different
algorithms. The first one is a direct deterministic algorithm, in
which one discretizes momentum space on a lattice in order to compute
the collision integrals by numerical quadratures (by assuming a
residual rotation invariance around the $p_z$ axis, we could perform
the azimuthal integrals analytically). The second method we have
considered is a variant of the direct simulation Monte-Carlo (DSMC),
in which the distribution is sampled by a large number of
``simulated'' particles.

The outcome of these computations is that the classical approximation
is not always guaranteed to be good --even at a qualitative level-- in
situations where the occupation number is large. At moderate
values of the coupling constant, the classical attractor scenario
cartooned in Figure \ref{fig:iso} is never observed, and the pressure
ratio $P_{_L}/P_{_T}$ becomes constant at late times or even increases
without showing any sign of a $\tau^{-2/3}$ behavior in the full
calculation (while a $\tau^{-2/3}$ behavior is indeed seen at late
times in the classical approximation). Increasing the occupation
number at fixed coupling does not make the classical approximation any
better.  The classical behavior, with $\tau^{-2/3}$ behavior in the
longitudinal pressure, does emerge when one increases $f_0$ and
decreases the coupling $g^2$, keeping $g^2 f_0$ constant.  But even in
this case, the full quantum behavior deviated from the classical one
when the occupancy at $p\sim Q$, $p_z=0$ was still large.

This study indicates that the conventional criterion for classicality,
$f\gg 1$, is too simple in situations where $f$ has a strong momentum
dependence. If this condition is meant to be understood as $f(p)\gg 1$
for all $p$, then it is not useful, because it is never realized. If
instead one understands ``$f$'' as the maximal value of $f(p)$, then
this condition is necessary for the classical approximation, but by no
means sufficient. Given this, the outcome of computations done in this
approximation should be considered with caution unless confirmed by
other computations performed in a framework that goes beyond this
classical approximation. In addition, extrapolations of classical
calculations from very weak coupling (where the classical and full
calculations agree over some extended time window) to larger couplings
must be taken with care, since the classical attractor behavior completely
disappears at couplings that are still relatively small.

The limitation we have discussed in this paper is due to the missing
$f^2$ terms when one uses the classical approximation in the collision
term of the Boltzmann equation. Therefore, it does not affect the
variant of the classical approximation mentioned in the last paragraph
of Section \ref{sec:boltz}, since the replacement $f\to
f+\tfrac{1}{2}$ that one performs in this variant restores the $f^2$
terms. However, this variant suffers from a potentially severe
sensitivity to the ultraviolet cutoff. For a non-expanding system, one
can mitigate this problem by choosing the cutoff a few times above the
physical scale. In the appendix \ref{sec:Non-expanding system}, we
show that this variant of the classical approximation describes
isotropization in a non-expanding system much better than the plain
classical approximation that has only the $f^3$ terms, without being
much affected by the ultraviolet cutoff. The cutoff dependence of
this variant of the classical approximation is much harder to keep
under control in the expanding case, because the physical scales (and
possibly the cutoff itself, depending on the details of the
implementation) are time dependent.  It also has much more severe
problems when the collision term includes number-changing processes.

Let us end by mentioning the very recent work presented in
ref.~\cite{KurkeZ1} where a similar study has been performed in the
case of Yang-Mills theory, by applying the effective kinetic theory of
ref.~\cite{ArnolMY5} to the study of a longitudinally expanding system
of gluons.  In this work, the authors also observe a rapid separation
between the full evolution and the classical approximation, already
for small couplings such as $g^2 N_c=0.5$.

\section*{Acknowledgements}
We would like to thank J.-P.\ Blaizot, J.\ Liao, A.\ Kurkela, N.\
Tanji and V.\ Greco for useful discussions.  B.W. would also like to
thank Z.\ Xu for some useful discussions about the test particle
method. This research was supported by the Agence Nationale de la
Recherche project 11-BS04-015-01 and by the Natural Sciences and
Engineering Research Council of Canada (NSERC).  Part of the
computations were performed on the Guillimin supercomputer at McGill
University, managed by Calcul Qu\'ebec and Compute Canada. The
operation of this supercomputer is funded by the Canada Foundation for
Innovation (CFI), Minist\`ere de l'Economie, de l'Innovation et des
Exportations du Qu\'ebec (MEIE), RMGA and the Fonds de recherche du
Qu\'ebec - Nature et technologies (FRQ-NT).

\appendix

\section{Anisotropic system in a fixed volume}
\label{sec:Non-expanding system}
In this appendix, we consider an anisotropic system of scalar
particles in a fixed volume, in order to study how the classical
approximation affects its isotropization. In this case, assuming that
the particle distribution depends only on time but not on position,
the left-hand side of eq.~(\ref{eq:bolt-SK}) reduces to
\begin{align}
(p^\mu \partial_\mu)\; f(\pp,\pz)=\omega_{p}\partial_t\, f(\omega_{p},\pz)\,,
\end{align}
so that the Boltzmann equation now reads
\begin{align}
\partial_t\, f(\omega_p,\pz)=C_{_{\rm nc}}[f]\,,\label{eq:boltfinalcontaniso}
\end{align}
with the collision term as given in eq (\ref{eq:boltafdelta}). The
evaluation of the collision term is identical to the case of a
longitudinally expanding system, while the free streaming part of the equation
is now completely trivial.

To obtain the numerical results presented in this appendix, we use
again an initial condition of the form:
\begin{align}
  f_{{\rm init}}(\omega_p,\pz)=\null f_0\;\exp\left(-\alpha\,
    \tfrac{\omega_p^2}{Q^2}-\beta\, \tfrac{\pz^2}{Q^2}\right)\, ,
\label{eq:finitdics-1}
\end{align}
with $f_0=100$, $\alpha=2$ and $\beta=1.2$. Since the
system is not expanding, the value of the initial time is
irrelevant. We have taken $Q t_0=0.1$. The coupling constant is
$g^4=50$ and the mass of the particles is set to $m/Q=0.1$. The number
of lattice spacings are set to $N_{\rm f}=2N_z=64$, while the maximal
values of $\pz$ and $\omega_p$ are given by $L/Q=3$ and
$\omega_{_\Lambda}=\sqrt{L^2+m^2}$.

In the case of an homogeneous non-expanding system, the conservation
laws take a very simple form:
\begin{equation}
n=\mbox{constant}\,,\qquad \epsilon=\mbox{constant}\, ,
\end{equation}
that we use to monitor the accuracy of the numerical
solution. Starting from the same initial condition given in
eq.~(\ref{eq:finitdics-1}), we have studied the time evolution of the
system with the full collision term (\ie\ with both the cubic and
quadratic terms in $f$), in the classical approximation (with the
cubic terms only), and in the ``classical statistical approximation''
(CSA) that amounts to including the zero point vacuum fluctuations
in the corresponding classical field approximation. See Section
\ref{sec:class-approx} for more details on the different classical
schemes.

\begin{figure}[htbp]
\begin{center}
\resizebox*{9cm}{!}{\includegraphics{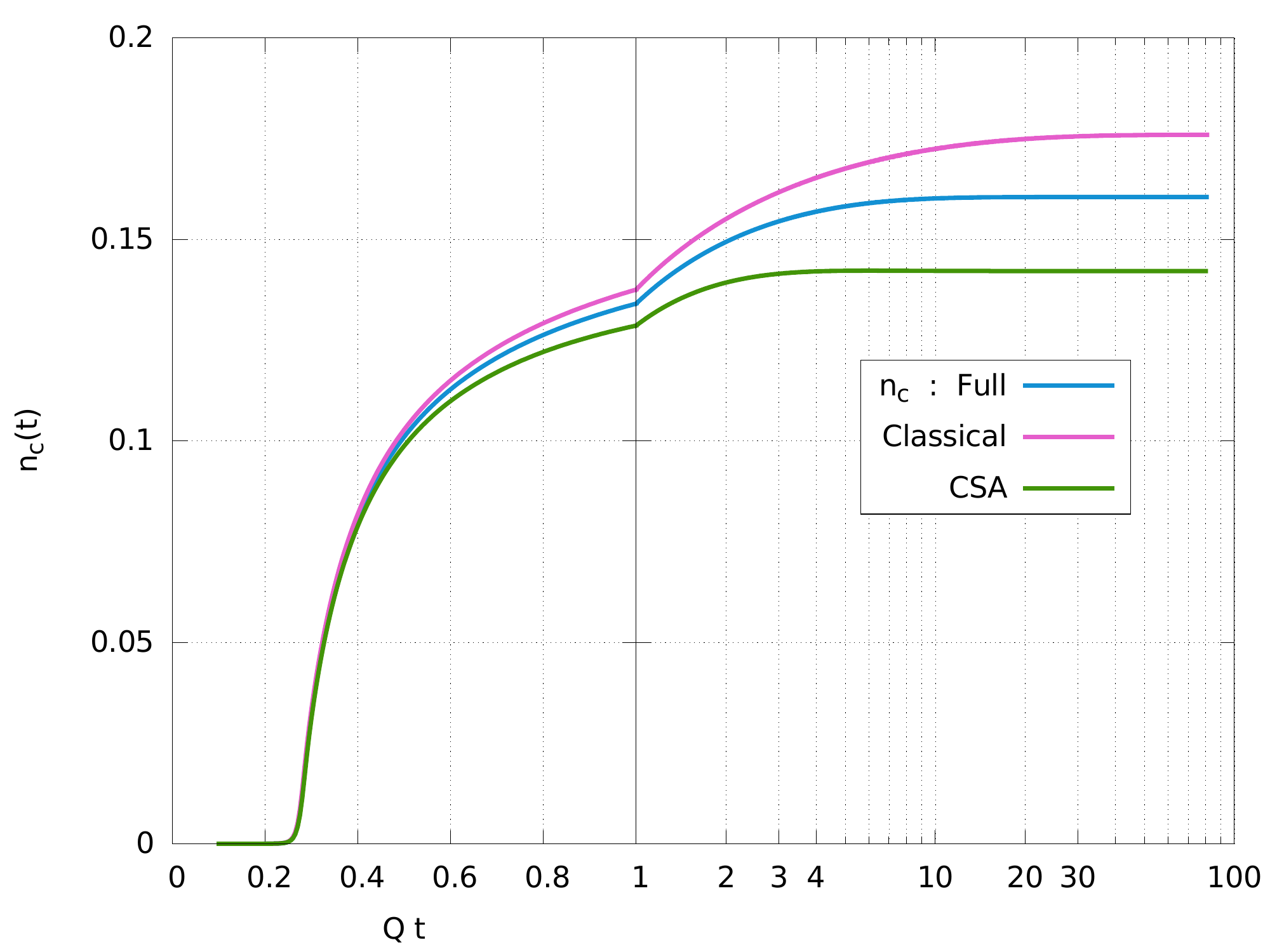}}
\end{center}
\caption{\label{fig:encondaniso2}Time evolution of the particle density in the condensate.}
\end{figure}

In Figure \ref{fig:encondaniso2}, we compare the time evolution of
the particle density in the condensate, $n_c$, for the three schemes.
It appears that the three schemes agree qualitatively, and even
semi-quantitatively, for the evolution of this quantity. The onset of
condensation is almost exactly identical in all the schemes, while the
final values of $n_c$ differ by about $10$\%. In agreement with the
observations of ref.~\cite{EpelbGTW1}, the classical approximation
tends to overpredict the fraction of condensed particles, while the
classical statistical approximation tends to underpredict it.

\begin{figure}[htbp]
\begin{center}
\resizebox*{9cm}{!}{\includegraphics{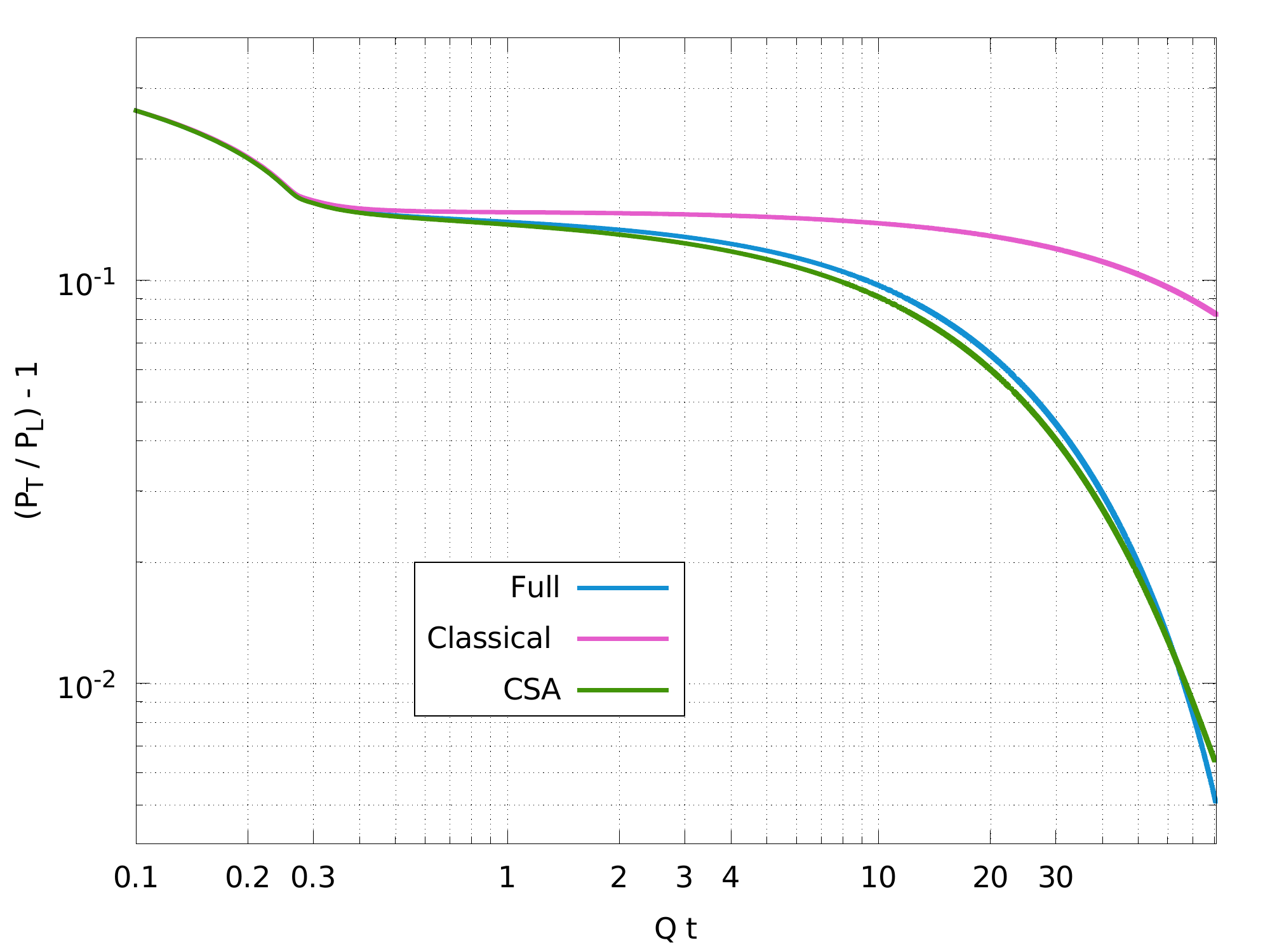}}
\end{center}
\caption{\label{fig:ratioaniso}Time evolution of ${P_{_T}}/{P_{_L}}-1$.}
\end{figure}

Next, we consider the time evolution of the ratio between the
transverse and longitudinal pressures. In Figure
\ref{fig:ratioaniso}, we plot the time evolution of
${P_{_T}}/{P_{_L}}-1$ in the three schemes. Starting from a nonzero
value dictated by the momentum anisotropy of the initial condition,
this quantity is expected to return to zero as the particle
distribution isotropizes.
After a short initial stage during which the three schemes are nearly
undistinguishable, we observe that the unapproximated scheme and the
classical statistical approximation lead to almost identical time
evolutions for this quantity, while the classical approximation
isotropizes at a much slower pace. This is consistent with the
argument exposed in the introduction, according to which the terms
quadratic in $f$ are essential for out-of-plane scatterings in an
anisotropic system.

\section{Integral of the product of four Bessel functions}
\label{sec:apa}
\subsection{Expression as an elliptic integral}
In eq (\ref{eq:boltbefdelta}), we have encountered an integral involving
the product of four Bessel $J_0$ functions,
\begin{align}
{\bs I}_4(\{p_i\})=\null&
\int_0^{+\infty}r\,\d r \,J_0(p_1 r) \, J_0(p_2 r) \,J_0(p_3 r) \,J_0(p_4 r)\,,\label{eq:pb}
\end{align}
which, to the best of our knowledge, does not seem to be known in the
literature. In this appendix, we derive an exact expression of this
integral in terms of an elliptic function $K$, starting from the known
expressions of similar integrals with two  $J_0$ Bessel functions (see
ref.~\cite{GradsR1}-6.512.8):
\begin{equation}
{\bs I}_2(\{p_i\})=\int_0^{+\infty}r\,\d r \,J_0(p_1 r) \, J_0(p_2 r)=
\frac{1}{p_1}\,\delta(p_1-p_2)\, ,
\label{eq:int2bessel}
\end{equation}
and three $J_0$ Bessel functions (see ref.~\cite{GervoN1}):
\begin{equation}
{\bs I}_3(\{p_i\})=\int_0^{+\infty}r\,\d r \,J_0(p_1 r) \, J_0(p_2 r) \,J_0(p_3 r) =
\frac{1}{2\pi{\cal A}(p_1,p_2,p_3)}\, .
\label{eq:int3bessel}
\end{equation}
${\cal A}(p_1,p_2,p_3)$ is the area of the triangle whose edges have
lengths $p_1,p_2$ and $p_3$ (if such a triangle does not exist, then
the integral is zero). We can therefore recast eq.~(\ref{eq:pb}) into
the following expression:
\begin{align}
{\bs I}_4(\{p_i\})=\null& \int\limits_0^{+\infty}rs \,\d r\, \d s\, \frac{\delta(r-s)}{s}
 J_0(p_1 r)  \, J_0(p_2 r)\, J_0(p_3 s) \,J_0(p_4 s)\notag\\
=\null&\int\limits_0^{+\infty}\!t\d t
 \!\int\limits_0^{+\infty}\!r\d r \, J_0(p_1 r) \, J_0(t r) \, J_0(p_2 r)
\!\int\limits_0^{+\infty}\!s\d s \, J_0(p_3 s)  \,J_0(t s) \,J_0(p_4 s)\notag\\
=\null&
\frac{1}{4\pi^2}\int_0^{+\infty}t \,\d t\,\frac{1}{{\cal A}(p_1,p_2,t)}
\,
\frac{1}{{\cal A}(p_3,p_4,t)}\,.
\end{align}
Recall that the area of a triangle in terms of the lengths of its
edges is given by the following formula,
\begin{eqnarray}
{\cal A}(a,b,c)&\smeq&\frac{1}{4} \sqrt{(a+b+c)(a+b-c)(a+c-b)(b+c-a)}
\nonumber\\
&\smeq&\frac{1}{4}\sqrt{((a+b)^2-c^2)(c^2-(a-b)^2)}\, .
\end{eqnarray}
Thus, eq.~(\ref{eq:pb}) can be expressed as
\begin{equation}
{\bs I}_4(\{p_i\})
=
\frac{2}{\pi^2}\int_0^\infty \frac{\d x}{
\sqrt{(\alpha_{12}-x)(x-\beta_{12})}
\sqrt{(\alpha_{34}-x)(x-\beta_{34})}
}\, ,
\label{eq:F-ellip}
\end{equation}
where we denote $\alpha_{ij}\equiv (p_i+p_j)^2$ and
$\beta_{ij}\equiv(p_i-p_j)^2$. The new integration variable is
$x\equiv t^2$. The range of integration on $x$ is restricted by the fact
that the arguments of the two square roots must both be positive:
\begin{eqnarray}
\beta_{12}< x< \alpha_{12}\,,\qquad \beta_{34}<x<\alpha_{34}\, .
\label{eq:cond}
\end{eqnarray}
Since it involves only a square root of a fourth degree polynomial,
the integral in eq.~(\ref{eq:F-ellip}) is an elliptic integral, which
can be reduced to a combination of Legendre's elliptic functions.

\subsection{Expression in terms of the elliptic $K$ function}
\begin{figure}[htbp]
\begin{center}
\resizebox*{8cm}{!}{\includegraphics{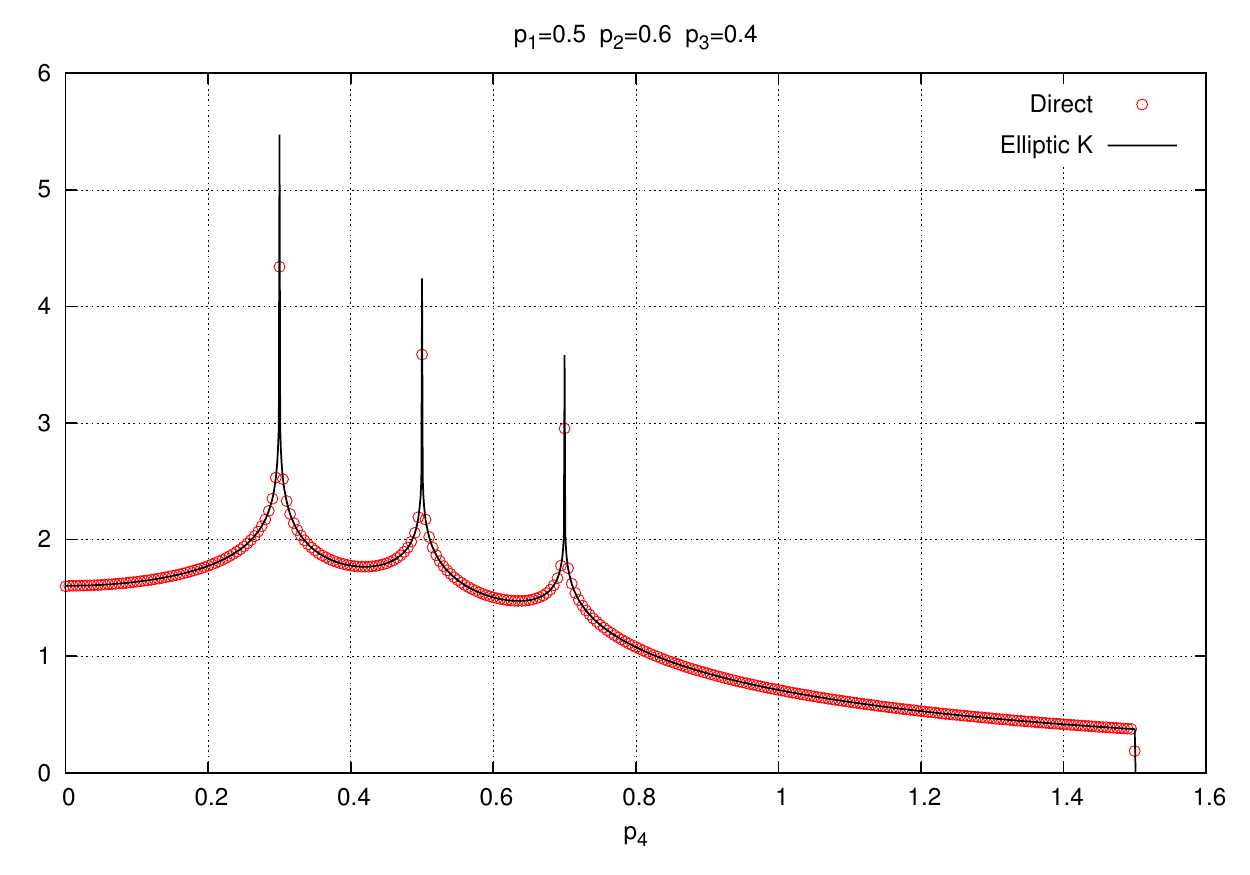}}
\end{center}
\caption{\label{fig:F}${\bs I}_4(\ppa{i})$ as a function of
  $\ppa{4}$ for fixed values of $\ppa{1},\ppa{2}$ and
  $\ppa{3}$. The dots represent the direct numerical computation of
  the integral in eq.~(\ref{eq:pb}), and the solid line is based on
  eqs.~(\ref{eq:F-ellip3}) and (\ref{eq:AGM}).}
\end{figure}

The boundaries of the integration range in eq.~(\ref{eq:F-ellip}) are
two of the roots of the polynomial
\begin{equation}
f(x)\equiv (\alpha_{12}-x)(x-\beta_{12})(\alpha_{34}-x)(x-\beta_{34})\, .
\end{equation}
Let us call $r_1$ and $r_2$ these two roots, respectively the lower
and upper bound. And for definiteness, let us call $r_3$ and $r_4$ the
remaining two roots, arranged so that $r_3<r_1<r_2<r_4$, \ie\
\begin{eqnarray}
&&
r_3\equiv\min(\beta_{12},\beta_{34})\quad,\qquad
r_1\equiv\max(\beta_{12},\beta_{34})\,,\nonumber\\
&&
r_2\equiv\min(\alpha_{12},\alpha_{34})\quad,\qquad
r_4\equiv\max(\alpha_{12},\alpha_{34})\,.
\end{eqnarray}
Eq.~(\ref{eq:pb}) is thus equivalent to
\begin{equation}
{\bs I}_4(\{p_i\})
=
\frac{2}{\pi^2}\int_{r_1}^{r_2} \frac{\d x}{
\sqrt{(r_2-x)(x-r_1)(r_4-x)(x-r_3)}
}\, .
\label{eq:F-ellip1}
\end{equation}
Let us now perform the following change of variables:
\begin{equation}
x=r_1\cos^2\theta+r_2\sin^2\theta\, .
\end{equation}
The above integral becomes
\begin{equation}
{\bs I}_4(\{p_i\})
=
\frac{4}{\pi^2(r_2-r_1)}\int_{0}^{\pi/2} \frac{\d\theta}{
\sqrt{(\alpha^2+\cos^2\theta)(\beta^2+\sin^2\theta)}
}\, ,
\label{eq:F-ellip2}
\end{equation}
with $\alpha^2\equiv(r_4-r_2)/(r_2-r_1)>0$ and
$\beta^2\equiv(r_1-r_3)/(r_2-r_1)>0$.  This integral can be expressed
in terms of the complete Legendre elliptic function of the first
kind (defined for $z\in[0,1)$),
\begin{equation}
{K}(z)\equiv\int_0^{\pi/2}\frac{\d\theta}{\sqrt{1-z\sin^2\theta}}\, ,
\end{equation}
which leads to the following compact expression\footnote{From this
  formula, one can check the identity ${\bs
    I}_4(p_1,p_2,p_3,0)=(2\pi{\cal A}(p_1,p_2,p_3))^{-1}$, that one
  expects from eqs.~(\ref{eq:pb}) and (\ref{eq:int3bessel}).}
\begin{equation}
{\bs I}_4(\{p_i\})
=
\frac{4\;K\!\left({\frac{(r_2-r_1)(r_4-r_3)}{(r_2-r_3)(r_4-r_1)}}\right)}
{\pi^2\,\sqrt{(r_2-r_3)(r_4-r_1)}}
\, .
\label{eq:F-ellip3}
\end{equation}
The function $K(z)$ can be evaluated efficiently with a simple algorithm
based on the arithmetic-geometric mean:
\begin{eqnarray}
&&
u_0=1\,,\; v_0=\sqrt{1-z}
\qquad u_{n+1}=\frac{u_n+v_n}{2}\,,\;v_{n+1}=\sqrt{u_nv_n}\nonumber\\
&&
K(z)=\frac{\pi}{u_{\infty}+v_{\infty}}\, .
\label{eq:AGM}
\end{eqnarray}
Figure \ref{fig:F} shows a comparison of a direct numerical
evaluation of the integral in eq.~(\ref{eq:pb}) with the formula
(\ref{eq:F-ellip3}).
Note that this quantity becomes singular for special configurations of
the $\ppa{i}$'s (in particular, when their values allow the vectors to
become collinear). Near these points, the direct method is
inefficient because of the very slow convergence of the integral.  In
contrast, eq.~(\ref{eq:F-ellip3}) is much better because the
algorithm described in eqs.~(\ref{eq:AGM}) converges to a very
accurate result in only a few iterations\footnote{The number of
  accurate digits doubles at every iteration.}. Moreover, this method
does not require the evaluation of any transcendental function.

\section{Integration domain for the collision term}
\label{sec:apb}
\begin{figure}
\begin{center}
\resizebox*{7cm}{!}{\includegraphics{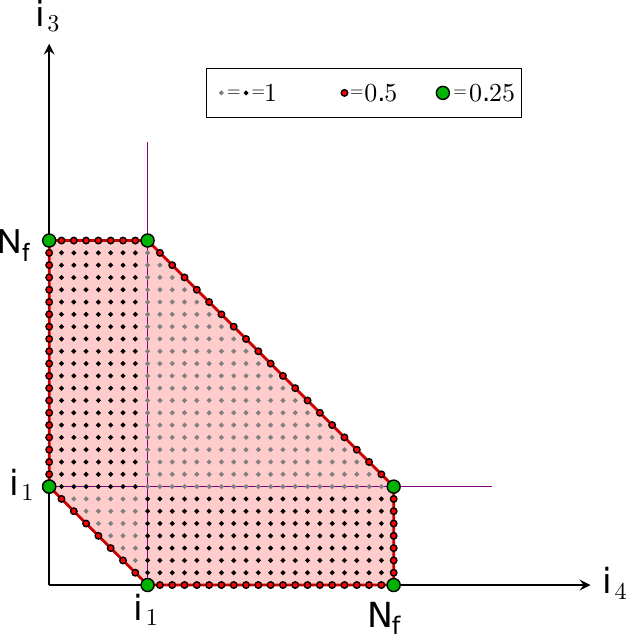}}
\end{center}
\caption{\label{dig:dom-om34}Discretization of the integrals over
  $\omega_{p_3}$ and $\omega_{p_4}$, with the values of the quadrature
  weights of each point.}
\end{figure}

\begin{figure}
\begin{center}
\resizebox*{7cm}{!}{\includegraphics{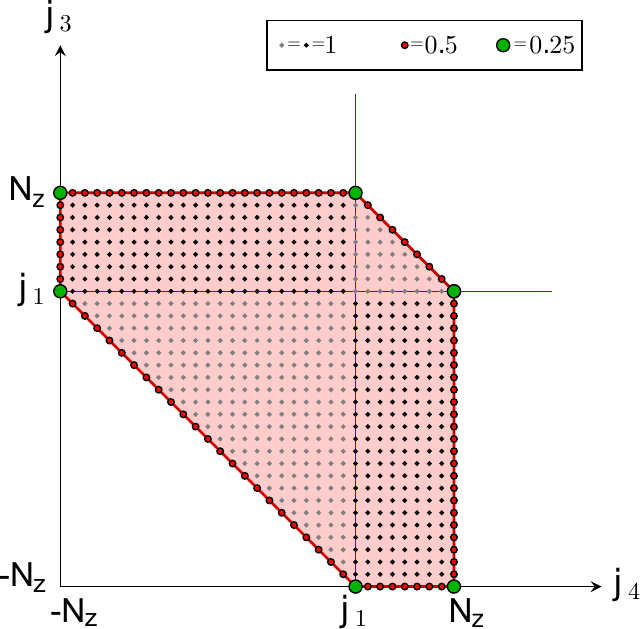}}
\end{center}
\caption{\label{dig:dom-pz34}Discretization of the integrals over
  $\pza{3}$ and $\pza{4}$, with the values of the quadrature
  weights of each point.}
\end{figure}

When we discretize the collision integral of
eq.~(\ref{eq:boltafdelta}), the domain of integration for the energy
variables $\omega_{p_3}$ and $\omega_{p_4}$ is the one represented in
Figure \ref{dig:dom-om34}.
When $i_1=1$ or $i_1=N_{\rm f}$, this domain becomes a triangle. In
this case, the $6$ points represented in green merge pairwise to form
the summits of the triangle. The quadrature weight at these points
becomes $\tfrac{1}{2}\times \tfrac{1}{2}\times
\tfrac{1}{2}=\tfrac{1}{8}$. But we do not need to handle this case by
hand since the formula (\ref{eq:discint}) gives the correct weights in
all cases. Likewise, we have represented the integration domain on the
longitudinal momenta $\pza{3}$ and $\pza{4}$ in Figure
\ref{dig:dom-pz34}, with the quadrature weight of each point.
Since we have assumed that $\pza{1}>0$, the index $j_1$ is positive.

We need also to specify the integration domains for the terms
$C^{_{1c\leftrightarrow 34}}_{p_1}[f]$ and $C^{_{12\leftrightarrow
    c4}}_{p_1}[f]$ that describe collisions between a particle from
the condensate and a particle at non-zero momentum, whose expressions
are given in eq.~(\ref{eq:Cc34}). When $\pza{1}>0$, the integration
over $\omega_{p_4}$ in $C^{_{1c\leftrightarrow 34}}_{p_1}[f]$ is
discretized as a sum over the following discrete points:
\setbox2\hbox to
4cm{\resizebox*{4cm}{!}{\includegraphics{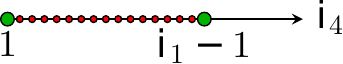}}}
\begin{equation}
\raise-4.5mm\box2
\qquad
i_4\in[1,i_1-1]\,,
\end{equation}
while for the longitudinal momentum we have:
\setbox2\hbox to 5.6cm{\resizebox*{5.6cm}{!}{\includegraphics{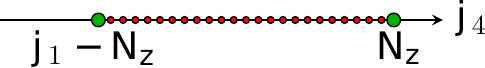}}}
\begin{equation}
\raise-4.5mm\box2
\qquad
j_4\in[j_1-N_z,N_z]\,.
\end{equation}
For $C^{_{12\leftrightarrow c4}}_{p_1}[f]$ given in
eq.~(\ref{eq:Cc34}), the integration domain for the energy
$\omega_{p_4}$ is \setbox3\hbox to
4cm{\resizebox*{4cm}{!}{\includegraphics{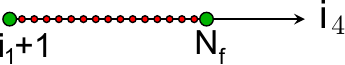}}}
\begin{equation}
\raise-4.5mm\box3
\qquad
i_4\in[i_1+1,N_{\rm f}]\,,
\end{equation}
while for the longitudinal momentum $\pza{4}$ we have
\setbox4\hbox to 5.6cm{\resizebox*{5.6cm}{!}{\includegraphics{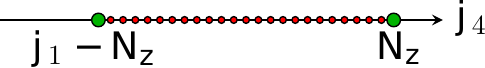}}}
\begin{equation}
\raise-4.5mm\box4
\qquad
j_4\in[j_1-N_z,N_z]\,.
\end{equation}

Finally, we need also to specify the discrete domains in the right
hand side of the equation (\ref{eq:Cc1}) for the evolution of the
particle density in the condensate. The sum over the energies
$\omega_{p_3}$ and $\omega_{p_4}$ is over the following set of
points:\setbox4\hbox to
6cm{\resizebox*{6cm}{!}{\includegraphics{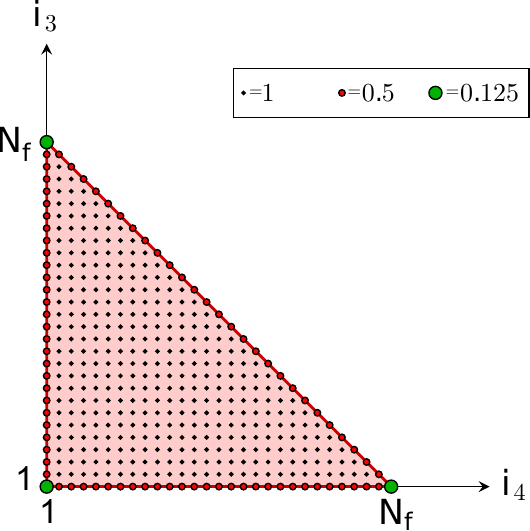}}}
\begin{equation}
\raise 0mm\box4
\nonumber
\end{equation}
and for the longitudinal momenta $\pza{3}$ and $\pza{4}$ it reads:
\setbox4\hbox to 6cm{\resizebox*{6cm}{!}{\includegraphics{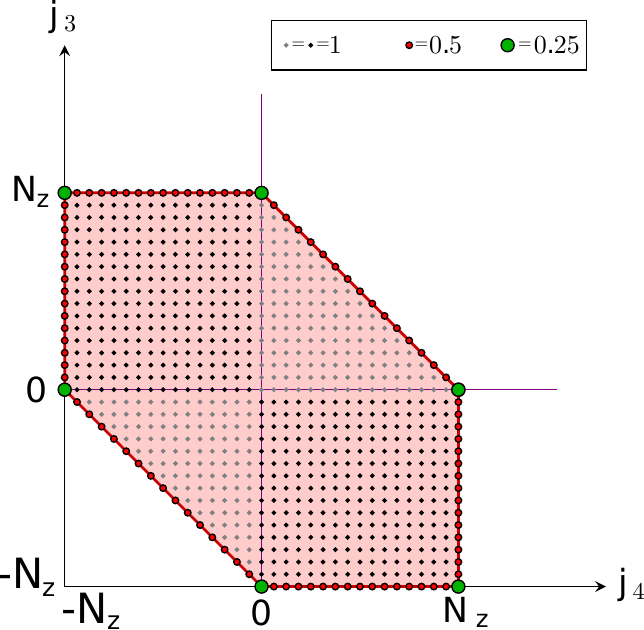}}}
\begin{equation}
\raise 0mm\box4\nonumber
\end{equation}

\section{Discretization of the free-streaming term}
\label{sec:apc}
In this section we derive the weights (\ref{eq:coeffsol}) that enter
into eq.~(\ref{eq:n1}). This derivation can be done in the case where
there is no Bose-Einstein condensate, since the particles in the BEC
carry zero momentum and are therefore not affected by free
streaming. Let us start with the particle density defined in
eq.~(\ref{eq:discpartnum}). From the conservation of the number of
particles, it should obey
\begin{align}
\tau\partial_\tau n_{\rm nc} =-n_{\rm nc}\,.
\end{align}
Let us recall that eq.~(\ref{eq:n1}) is only valid for $j>
0$. Therefore, we first need to rewrite the particle density
(\ref{eq:discpartnum}) by using the parity in $\pz$
(\ie\ $h[i,-j]=h[i,j]$) and the fact that $ w_z[-j]= w_z[j]$
\begin{align}
n_{\rm nc}=\sum_{i=1}^{N_{\rm f}}w_{\rm f}[i] w_z[0]h[i,0]
+2\sum_{i=1}^{N_{\rm f}}\sum_{j=1}^{N_z}w_{\rm f}[i]\, w_z[j]\;h[i,j]\,.
\label{eq:nnc1}
\end{align}
As one can see on Figure \ref{fig:freestrdisc}, particles can hop
to the $j=0$ line from the left or from the right. Therefore, for
$j=0$, eq.~(\ref{eq:n1}) can be rewritten as follows:
\begin{eqnarray}
\tau\partial_\tau \big(w_{\rm f}[i] w_z[0]\, h[i,0]\big)&\smeq&
{} -\alpha_{i0}\,w_{\rm f}[i]\, w_z[0]\, h[i,0]\nonumber\\
&&{} +2\beta_{i1}\,w_{\rm f}[i]\, w_z[1]\,h[i,1]\nonumber\\
&&{} +2\gamma_{i+1 1}\,w_{\rm f}[i+1]\, w_z[1]\, h[i+1,1] \label{eq:n10}\,.
\end{eqnarray}
Then , by summing eqs.~(\ref{eq:n1}) and (\ref{eq:n10}) on the indices
$i,j$, we obtain (the indices of the last two terms have been shifted)
\begin{align}
n_{\rm nc}\null&=
\sum_{i=1}^{N_{\rm f}}\alpha_{i0}w_{\rm f}[i] w_z[0]h[i,0]
+2\sum_{i=1}^{N_{\rm f}}\sum_{j=1}^{N_z}\alpha_{ij} w_{\rm f}[i] w_z[j]  h[i,j]\notag\\
&
-2\sum_{i=1}^{N_{\rm f}}\sum_{j=1}^{N_z}\beta_{ij} w_{\rm f}[i] w_z[j]  h[i,j]
-2\sum_{i=2}^{N_{\rm f}}\sum_{j=1}^{N_z}\gamma_{ij} w_{\rm f}[i] w_z[j]  h[i,j]\,.
\end{align}
The right-hand side must therefore be equal to that of
eq.~(\ref{eq:nnc1}), for every $h[i,j]$. This leads first to
\begin{align}
\alpha_{i 0}=\null&1\, ,
\end{align}
and
\begin{eqnarray}
&\alpha_{1j}-\beta_{1j}=1\qquad &\mbox{if\ } j>0\,,\nonumber\\
&\alpha_{ij}-\beta_{ij}-\gamma_{ij}=1\qquad &\mbox{if\ } i>1, j>0\,.\label{eq:rel1}
\end{eqnarray}
Next, using the definition for the total energy (\ref{eq:discener})
and the longitudinal pressure (\ref{eq:discPL}), we can use Bjorken's
law
\begin{align}
\tau\partial_\tau \epsilon_{\rm nc}=\null& -\epsilon_{\rm nc}-P_{_L{\rm nc}}\,,
\end{align}
in order to obtain
\begin{align}
\epsilon_{\rm nc}+P_{_L{\rm nc}}\null&=\sum_{i=1}^{N_{\rm f}}w_{\rm f}[i] w_z[0]\alpha_{i0} \epsilon[i,0]
+2\sum_{i=1}^{N_{\rm f}}\sum_{j=1}^{N_z}w_{\rm f}[i] w_z[j]\alpha_{ij} \epsilon[i,j]\notag\\
&-2\sum_{i=1}^{N_{\rm f}}\sum_{j=1}^{N_z} w_{\rm f}[i] w_z[j]\beta_{ij} \epsilon[i,j]\notag\\
&-2\sum_{i=2}^{N_{\rm f}-1}\sum_{j=1}^{N_z} w_{\rm f}[i] w_z[j]
\gamma_{ij} \frac{\omega_p[i-1]}{\omega_p[i]} \epsilon[i,j]\,,
\end{align}
where we have denoted $\epsilon[i,j]\equiv\omega_p[i]h[i,j]$. This
implies the following constraints among the coefficients
$\alpha_{ij},\beta_{ij},\gamma_{ij}$:
\begin{eqnarray}
\alpha_{i0}&\smeq&1 \nonumber\\
\alpha_{1j}-\beta_{1j}
   &\smeq&1 + \frac{\pz^2[j]}{\omega_p^2[1]} \qquad \mbox{if\ } i=1, j>0\,,\nonumber\\
\alpha_{ij}-\beta_{ij}-\frac{\omega_p[i-1]}{\omega_p[i]}\gamma_{ij}
  &\smeq&1 + \frac{\pz^2[j]}{\omega_p^2[1]}
     \qquad \mbox{if\ } i>1, j>0\,.\label{eq:rel1-1}
\end{eqnarray}
The second of these constraints is incompatible with the first of
eqs.~(\ref{eq:rel1}), unless we set up the lattice spacings in such a
way that the points $(1,j>0)$ do not satisfy the mass-shell condition
(\ie\ are below the red line in Figure \ref{fig:freestrdisc}), so
that they do not contribute to $n,\epsilon$ and $P_{_L}$.  From now
on, we assume that this is the case. We do not explicitly exclude
these points from the sums, but we simply assume that $h[1,j>0]=0$.

By comparing the second of eqs.~(\ref{eq:rel1}) and the third of
eqs.~(\ref{eq:rel1-1}), we obtain
\begin{equation}
\gamma_{ij}=\frac{\pz^2[j]}{\omega_p[i]\Delta\omega}\qquad\mbox{if\ } i>1, j>0\,. \label{eq:rel2}
\end{equation}
In order to fully constrain the coefficients, we need to consider also
the total longitudinal momentum given in
eq.~(\ref{eq:discfirstmoment}), and impose its conservation:
\begin{align}
\tau\partial_\tau (\rho_z)=-2\rho_z\,.
\end{align}
This leads to
\begin{align}
2\rho_z\null&=2\sum_{i=1}^{N_{\rm f}}\sum_{j=1}^{N_z} w_{\rm f}[i] w_z[j] \alpha_{ij}  \tilde{p}_z[i,j]
-2\sum_{i=1}^{N_{\rm f}}\sum_{j=1}^{N_z} w_{\rm f}[i] w_z[j]\frac{\pz[j-1]}{\pz[j]}\beta_{ij} \tilde{p}_z[i,j]\notag\\
&-2\sum_{i=2}^{N_{\rm f}}\sum_{j=1}^{N_z}
w_{\rm f}[i] w_z[j] \gamma_{ij} \frac{\pz[j-1]}{\pz[j]} \tilde{p}_z[i,j]\,,
\end{align}
where we denote $\tilde{p}_z[i,j]\equiv\pz[j]h[i,j]$. From this
equation, we obtain an additional constraint
\begin{align}
\alpha_{ij}-\frac{\pz[j-1]}{\pz[j]}\left(\beta_{ij}
+\gamma_{ij}\right)=\null&2\,.
\end{align}
Combining it with eqs.~(\ref{eq:rel1}) and (\ref{eq:rel2}), we get finally
\begin{eqnarray}
\beta_{ij}&\smeq&\frac{\pz[j]}{\Delta p_z}-\frac{\pz^2[j]}{\omega_p[i]\Delta\omega}\,,\\
\alpha_{ij}&\smeq&1+\frac{\pz[j]}{\Delta p_z}\,.\label{eq:rel3}
\end{eqnarray}

\section{Direct simulation Monte-Carlo method}
\label{app:dsmc}
In this appendix, we present a generalization of the so-called direct
simulation Monte Carlo (DSMC) method \cite{GarciW1} to study the
time-evolution of the distribution of relativistic particles and the
formation of a Bose-Einstein condensate in a boost-invariant
longitudinally expanding system. By including only the
$2\leftrightarrow2$ processes, the Boltzmann equation takes the
following general form,
\begin{eqnarray}
\left[\partial_\tau-\frac{p_{z_1}}{\tau}
\frac{\partial}{\partial p_{z_1}}\right]f(\p_{\perp1},p_{z_1})=C_{p_1}[f],
\end{eqnarray}
where the collision integral reads
\begin{eqnarray}
C_{p_1}[f]&\equiv&\frac{1}{4\omega_{p_1}}\int\limits_{\p_{2,3,4}}(2\pi)^4\delta^{(4)}(p_1+p_2-p_3-p_4)\nonumber\\
&&\qquad\qquad\qquad
\times\, |M(p_1,p_2;p_3,p_4)|^2\; F_{nc}({P_i})\, .
\end{eqnarray}
$M(p_1,p_2;p_3,p_4)$ is the matrix element for the scattering process
$p_1,p_2\leftrightarrow p_3,p_4$, and it simply reads
$M(p_1,p_2;p_3,p_4)=g^2$ for the $\phi^4$ theory.  Let us parameterize
the 3-momenta of the final state particles as follows:
\begin{eqnarray}
\p_3=\frac{1}{2}[\P_{\rm tot}+p\,{\bs\Omega}]\,,\qquad
\p_4=\frac{1}{2}[\P_{\rm tot}-p\,{\bs\Omega}]\,,\label{eq:outgoing}
\end{eqnarray}
where ${\bs\Omega}$ is a unit vector (${\bs\Omega}^2=1$) and $\P_{\rm
  tot}$ the total momentum
\begin{equation}
\P_{\rm tot}\equiv \p_1+\p_2\,.
\end{equation}
The conservation of energy gives
\begin{eqnarray}
p=|\p_3 - \p_4|
=\frac{E_{\rm tot}\sqrt{s-4 m^2}}{\sqrt{E_{\rm tot}^2-(\P_{\rm tot}\cdot{\bs \Omega}})^2}
 \,.\label{eq:a}
\end{eqnarray}
with $E_{\rm tot}\equiv \omega_{p_1}+\omega_{p_2}$ and $s\equiv
E_{tot}^2-\P_{tot}^2$.  By replacing $\p_4$ by the new variables
${\bs\Omega}$ and $p$ and integrating out $p$ and $\p_3$, the
Boltzmann equation can be shown to take the following form,
\begin{eqnarray}\label{eq:bolbInv}
&&\left(\f{\partial}{\partial \tau}
- \f{p_{z1}}{\tau}\f{\partial}{\partial p_{z1}}\right) f(\p_{\perp1},p_{z_1})
=
\int \frac{\d^3\p_2}{(2\pi)^3}
\int \frac{\d{\bs\Omega}}{4\pi}\;
 \frac{|M(p_1,p_2;p_3,p_4)|^2}{64\pi\, \omega_{p_1} \omega_{p_2}} \nonumber\\
&&
\qquad\qquad\qquad\qquad\qquad
\times\frac{E_{\rm tot}^2\sqrt{s-4m^2}}{[ E_{\rm tot}^2 - (\P_{\rm tot}\cdot{\bs\Omega})^2]^\frac{3}{2}}F_{\rm nc}(\{P_i\})\, ,
\end{eqnarray}
where $F_{\rm nc}(\{P_i\})$ is defined in eq.~(\ref{eq:Fnc}).

Eq.~(\ref{eq:bolbInv}) can be solved by the direct simulation
Monte-Carlo method. In this method, one defines a Markov process with
a total number $N$ of simulated particles~\cite{GarciW1}. Let us
introduce a partition $\{V_l\}$ (with $l\in[1, \cdots, M]$) of the
momentum space into $M$ bins and denote $N_l$ the number of simulated
particles with momentum in the momentum bin $V_l$. The distribution
function is then approximated by
\begin{eqnarray}\label{eq:fapprox}
\frac{V_l}{(2\pi)^3}\;f({p\in V_l})  \approx n\; \f{N_{l}}{N}\, ,
\end{eqnarray}
where $n$ is the number density of ``real'' particles. In this paper,
we assume that the system is homogeneous and isotropic in the
transverse plane. Therefore, we adopt a partition of momentum space
that divides the transverse momentum squared in equal intervals,
$0\le p_\perp^2\le p_{\perp{\rm max}}^2$, and likewise
for the longitudinal momentum axis, $0\le p_z\le p_{z,{\rm max}}$,
\ie
\begin{eqnarray}
  &&p_{\perp i}^2=(i+1)\, \Delta p_\perp^2\quad(0\le i < M_\perp)
  \qquad\text{with\ \ }
  \Delta p_\perp^2=\frac{p_{\perp {\rm max}}^2}{M_\perp}\, ,
\nonumber\\
&&p_{z j}=(j+1)\, \Delta p_z\quad(1\le j< M_z)
\qquad\text{with\ \ } \Delta p_z=\frac{p_{z,{\rm max}}}{M_z}\, .
\end{eqnarray}

Let us denote the momentum of the $s$-th simulated particle by $\p_s$
with $1\leq s\leq N$.  The probability for any two of the simulated
particles { $s_1$ and $s_2$ (with $1\leq s_1\neq s_2\leq
  N$) to scatter off each other during a time interval
\begin{equation}
\Delta t \equiv \f{2}{ n\, (N-1)\,\widehat{Y} }\label{eq:Dt}
\end{equation}
is given by
\begin{equation}
 P_{s_1 s_2}(\Omega) =
\frac{1}{4\pi}
\frac{2}{N(N-1)}
\frac{Y(\p_{s_1},\p_{s_2},{\bs\Omega})}{\widehat{Y}}\,,
\end{equation}
where
\begin{eqnarray}
Y(\p_{s_1},\p_{s_2},{\bs\Omega})
\equiv
&&\frac{E_{\rm tot}^2\sqrt{s-4m^2}\;|M(p_{s_1},p_{s_2};p_{s_1}',p_{s_2}')|^2}{64\pi \omega_{p_{s_1}} \omega_{p_{s_2}}[ E_{\rm tot}^2 - (\P_{\rm tot}\cdot{\bs\Omega})^2]^\frac{3}{2}}\nonumber\\
&&\qquad\qquad\quad\times\,[1+f(\p_{s_1}')+f(\p_{s_2}')]\, ,
\end{eqnarray}
with $E_{\rm tot}$, $\P_{\rm tot}$ respectively the sums of energies
and momenta of the particles $s_1$ and $s_2$ and $\p_{s_1}'$ and
$\p_{s_2}'$ given by eqs.~(\ref{eq:outgoing}) (with $p_1, p_2, p_3$
and $p_4$ respectively replaced by $p_{s_1},p_{s_2},p_{s_1}'$ and
$p_{s_2}'$)}. Here, $\widehat{Y}$ can be chosen arbitrarily {as long
as} it satisfies
\begin{eqnarray}
\widehat{Y}\geq Y(\p_{s_1},\p_{s_2},{\bs\Omega})\qquad\text{for all\ \ } s_1, s_2 \text{\ and\ \ } {\bs\Omega}\,.
\end{eqnarray}
For each time interval $\Delta t$, the effect of the longitudinal
expansion can be taken into account by simply rescaling the
longitudinal momenta of all the simulated particles, according to
\begin{equation}
\label{eq:pzRescale}
p_{z} \to \frac{t}{t+\Delta t}
p_{z}
 \end{equation}
 and
\begin{equation}
\label{eq:rhoRescale} n\to \frac{t}{t+\Delta t}\; n\,.
\end{equation}

In situations where a BEC may form, we define as belonging to the
condensate the simulated particles with momenta \footnote{This amounts
  to regularizing the delta function that would normally characterize
  the condensate by
\begin{eqnarray*}
\delta(\p)\quad
\to \quad
\frac{1}{2\pi p_{\perp {\rm min}}^2 p_{z,{\rm min}}}\;\theta(p_{\perp {\rm min}}-p_\perp)\,
\theta(p_{z, {\rm min}}-|p_z|)\, .
\end{eqnarray*}
This choice makes our algorithm faster than the one used in the test
particle method \cite{JacksZ1,XuZZG1} and hence better suited for the
study of expanding systems.  }
\begin{eqnarray}
p_\perp \le p_{\perp{\rm min}}<\Delta p_\perp\text{~~and~~}|p_z| \le p_{z,{\rm min}}<\Delta p_z\, ,
\label{eq:Ccond}
\end{eqnarray}
and the number density of the condensate  is defined as
\begin{eqnarray}
n_c^{_{\rm MC}}\equiv n\, \frac{N_{\rm c}}{N}
\end{eqnarray}
with $N_{\rm c}$ the number of condensate simulated particles that
satisfy the condition (\ref{eq:Ccond}). Note that this definition of
the particles in the condensate includes both particles with exactly
zero momentum (the ``genuine'' condensate), and the particles with
non-zero momentum in a small volume around $\p=0$. When this extra
volume is small enough, the above definition agrees well with an
actual condensate. {For all the results using DSMC in the
  expanding case have used the following parameters
\begin{eqnarray}
&&N=10^5\,,\qquad M_\perp=320\,,\qquad M_z=20\,,\nonumber\\
&&p_{\perp {\rm min}}=p_{z,{\rm min}}=Q/20\,,\qquad p_{\perp {\rm max}}=p_{z,{\rm max}}=6\,Q\,.
\end{eqnarray}
}

The DSMC algorithm for solving the Boltzmann equation is made of the
following steps:
\begin{enumerate}
\item[{\bf i.}]{Choose randomly a pair of simulated particles $(s_1, s_2)$
    (with a uniform probability distribution $\frac{2}{N(N-1)}$ among
    all the possible pairs)},
\item[{\bf ii.}]{Choose a random vector ${\bs\Omega}$ (with an uniform
    distribution on the unit sphere),}
\item[{\bf iii.}]{{Calculate $\p_{s_1}'$ and $\p_{s_2}'$
      from $\p_{s_1},\p_{s_2}$ and $\bs\Omega$.  If two or more
      momenta among $\p_{s_1}, \p_{s_2}, \p_{s_1}'$ and $\p_{s_2}'$
      satisfy eq.~(\ref{eq:Ccond}), skip the step {\bf iv} and go
      directly to {\bf v}.}}
\item[{\bf iv.}] Choose a random number $\xi \in [0,1]$ (with an
  uniform distribution).  If $\xi
  <\frac{Y(\p_{s_1},\p_{s_2},{\bs\Omega})}{\widehat Y}$,
  {$\p_{s_1}$ and $\p_{s_2}$ are respectively replaced by
    $\p_{s_1}'$ and $\p_{s_2}'$},
  \item[{\bf v.}]{Rescale the longitudinal momenta of all the
      simulated particles and the particle density $n$ according to
      eqs.~(\ref{eq:pzRescale}) and (\ref{eq:rhoRescale})},
  \item[{\bf vi.}]Increment the time $t\to t+\Delta t$ and return to
      the step {\bf i}.
\end{enumerate}

\begin{figure}[htbp]
\begin{center}
\resizebox*{9cm}{!}{\includegraphics{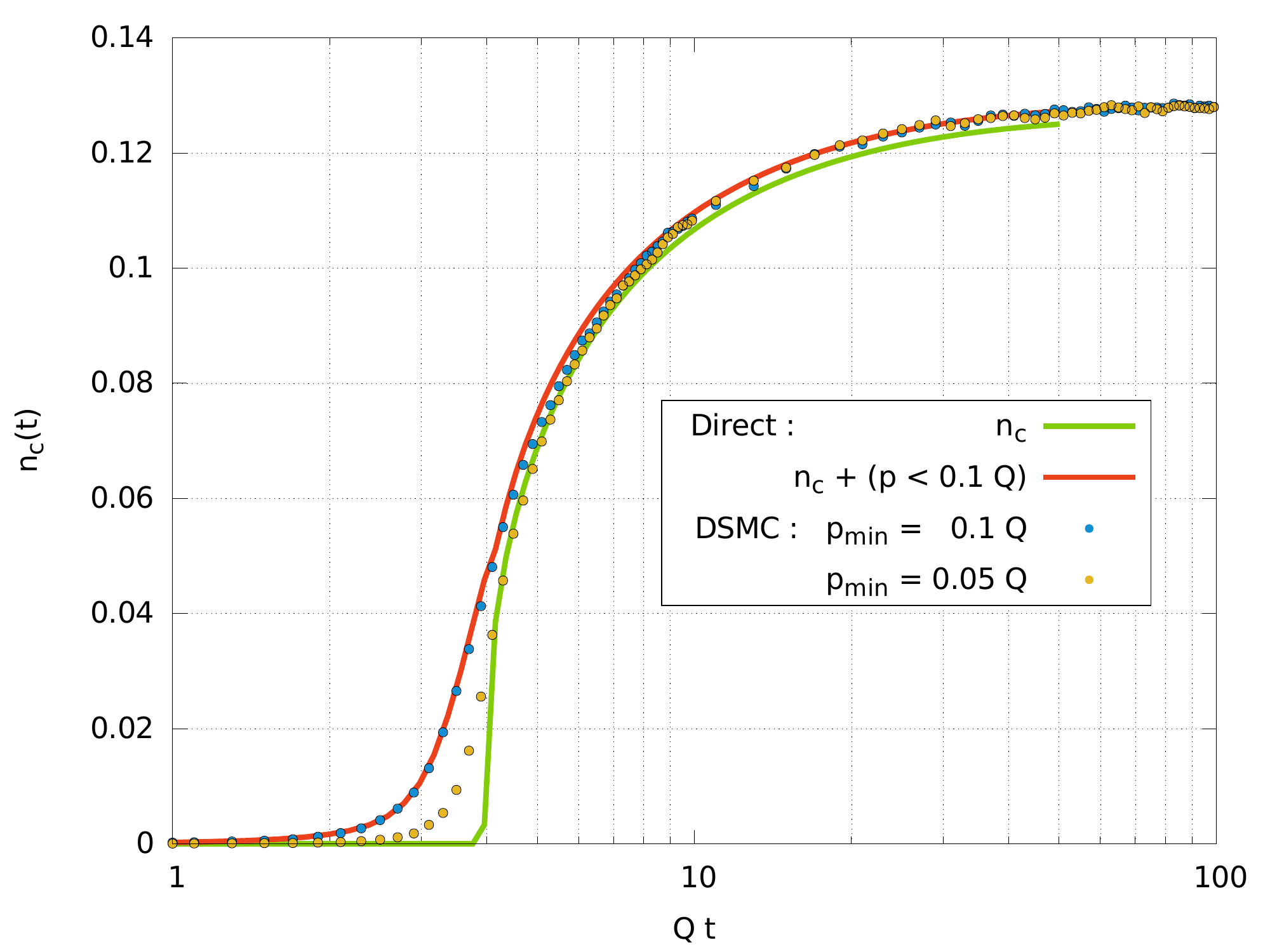}}
\end{center}
\caption{\label{fig:nccmp} Particle density in the condensate in the
  deterministic (``Direct'') and in the DSMC method. {Here, the
    distribution at $Qt=1$ is given by $f(Qt=1,p)=11.84\;\theta(Q-p)$.}  }
\end{figure}

In order to illustrate this method and its difference with the
deterministic method used in the rest of this paper, we have applied
it to the case of a {spatially homogeneous} non-expanding
system.  {In this case the Dirac function is regularized
  by $\delta(\p)\;\to\; \frac{3}{4\pi p_{\rm min}^3}\,\theta(p_{\rm min}-p)$
  and the momentum space is divided in equal intervals $\Delta p =
  0.2\,Q$ of the modulus of the momentum.}  The results of this
comparison are shown in Figure \ref{fig:nccmp}.
Because the operational definition of the condensate in the DSMC
method also includes the particles in a small sphere of radius $p_{\rm
  min}$ around $p=0$, the transition of condensation appears less
sharp than in the deterministic method where the density $n_{\rm c}$
is defined as the coefficient of the delta function $\delta(\p)$. By
decreasing the value of the $p_{\rm min}$ used in this definition, the
transition in the DSMC simulation appears to become sharper, to
finally become very close to the transition observed in the direct
method.  Alternatively, this interpretation can be further checked by
adding to the $n_{\rm c}$ of the deterministic method the contribution
of the non-condensed particles contained in this small sphere.


\begin{thebibliography}{10}

\bibitem{BergeBSV1}
{J. Berges, K. Boguslavski, S. Schlichting, R. Venugopalan}, Phys. Rev. {\bf D}
  {\bf 89}, 074011 (2014).

\bibitem{BergeBSV3}
{J. Berges, K. Boguslavski, S. Schlichting, R. Venugopalan}, JHEP {\bf 1405},
  054 (2014).

\bibitem{BergeBSV2}
{J. Berges, K. Boguslavski, S. Schlichting, R. Venugopalan}, Phys. Rev. {\bf D}
  {\bf 89}, 114007 (2014).

\bibitem{BlaizGLMV1}
{J.P. Blaizot, F. Gelis, J. Liao, L. McLerran, R. Venugopalan}, Nucl. Phys.
  {\bf A} {\bf 873}, 68 (2012).

\bibitem{DusliEGV2}
{K. Dusling, T. Epelbaum, F. Gelis, R. Venugopalan}, Phys. Rev. {\bf D} {\bf
  86}, 085040 (2012).

\bibitem{EpelbG3}
{T. Epelbaum, F. Gelis}, Phys. Rev. Lett. {\bf 111}, 232301 (2013).

\bibitem{KurkeM3}
{A. Kurkela, G.D. Moore}, Phys. Rev. {\bf D} {\bf 86}, 056008 (2012).

\bibitem{KurkeM1}
{A. Kurkela, G.D. Moore}, JHEP {\bf 1112}, 044 (2011).

\bibitem{KurkeL1}
{A. Kurkela, E. Lu}, Phys. Rev. Lett. {\bf 113}, 182301 (2014).

\bibitem{HelleMST1}
{M.P. Heller, D. Mateos, W. van der Schee, D. Trancanelli}, Phys. Rev. Lett.
  {\bf 108}, 191601 (2012).

\bibitem{HelleMST2}
{M.P. Heller, D. Mateos, W. van der Schee, M. Triana}, JHEP {\bf 1309}, 026
  (2013).

\bibitem{WuR1}
{B. Wu, P. Romatschke}, Int. J. Mod. Phys. {\bf C} {\bf 22}, 1317 (2011).

\bibitem{HelleJW1}
{M.P. Heller, R.A. Janik, P. Witaszczyk}, Phys. Rev. Lett. {\bf 108}, 201602
  (2012).

\bibitem{FukusG1}
{K. Fukushima, F. Gelis}, Nucl. Phys. {\bf A} {\bf 874}, 108 (2012).

\bibitem{ScheeRP1}
{W. van der Schee, P. Romatschke, S. Pratt}, Phys. Rev. Lett. {\bf 111}, 22,
  222302 (2013).

\bibitem{IancuLM3}
{E. Iancu, A. Leonidov, L.D. McLerran}, Lectures given at Cargese Summer School
  on QCD Perspectives on Hot and Dense Matter, Cargese, France, 6-18 Aug 2001,
  hep-ph/0202270.

\bibitem{IancuV1}
{E. Iancu, R. Venugopalan}, Quark Gluon Plasma 3, Eds. R.C. Hwa and X.N. Wang,
  World Scientific, hep-ph/0303204.

\bibitem{Weige2}
{H. Weigert}, Prog. Part. Nucl. Phys. {\bf 55}, 461 (2005).

\bibitem{GelisIJV1}
{F. Gelis, E. Iancu, J. Jalilian-Marian, R. Venugopalan}, Ann. Rev. Part. Nucl.
  Sci. {\bf 60}, 463 (2010).

\bibitem{Gelis15}
{F. Gelis}, Int. J. Mod. Phys. {\bf A} {\bf 28}, 1330001 (2013).

\bibitem{MuellS1}
{A.H. Mueller, D.T. Son}, Phys. Lett. {\bf B} {\bf 582}, 279 (2004).

\bibitem{Jeon3}
{S. Jeon}, Phys. Rev. {\bf C} {\bf 72}, 014907 (2005).

\bibitem{MathiMT1}
{V. Mathieu, A.H. Mueller, D.N. Triantafyllopoulos}, Eur. Phys. J. {\bf C} {\bf
  74}, 2873 (2014).

\bibitem{BergeBSV4}
{J. Berges, K. Boguslavski, S. Schlichting, R. Venugopalan}, Phys. Rev. Lett.
  {\bf 114}, 061601 (2015).

\bibitem{BergeSSV1}
{J. Berges, B. Schenke, S. Schlichting, R. Venugopalan}, Nucl. Phys. {\bf A}
  {\bf 931}, 348 (2014).

\bibitem{SemikT1}
{D.V. Semikoz, I.I. Tkachev}, Phys. Rev. Lett. {\bf 74}, 3093 (1995).

\bibitem{SemikT2}
{D.V. Semikoz, I.I. Tkachev}, Phys. Rev. {\bf D} {\bf 55}, 489 (1997).

\bibitem{EpelbG1}
{T. Epelbaum, F. Gelis}, Nucl. Phys. {\bf A} {\bf 872}, 210 (2011).

\bibitem{BlaizLM2}
{J.P. Blaizot, J. Liao, L.D. McLerran}, Nucl. Phys. {\bf A} {\bf 920}, 58
  (2013).

\bibitem{BlaizWY1}
{J.P. Blaizot, B. Wu, L. Yan}, arXiv:1402.5049.

\bibitem{BergeS4}
{J. Berges, D. Sexty}, Phys. Rev. Lett. {\bf 108}, 161601 (2012).

\bibitem{EpelbGTW1}
{T. Epelbaum, F. Gelis, N. Tanji, B. Wu}, Phys. Rev. {\bf D} {\bf 90}, 125032
  (2014).

\bibitem{EpelbGW1}
{T. Epelbaum, F. Gelis, B. Wu}, Phys. Rev. {\bf D} {\bf 90}, 065029 (2014).

\bibitem{Jeon1}
{S. Jeon}, Phys. Rev. {\bf D} {\bf 47}, 4586 (1993).

\bibitem{Jeon2}
{S. Jeon}, Phys. Rev. {\bf D} {\bf 52}, 3591 (1995).

\bibitem{ArnolMY6}
{P. Arnold, G.D. Moore, L.G. Yaffe}, JHEP {\bf 0011}, 001 (2000).

\bibitem{KurkeZ1}
{A. Kurkela, Y. Zhu}, arXiv:1506.06647.

\bibitem{ArnolMY5}
{P. Arnold, G.D. Moore, L.G. Yaffe}, JHEP {\bf 0301}, 030 (2003).

\bibitem{GradsR1}
{I.S. Gradshteyn, I.M. Ryzhik}, {\sl Table of integrals, series and products},
  Academic Press, London (2000).

\bibitem{GervoN1}
{A. Gervois, H. Navelet}, J. Math. Phys. {\bf 25}, 3350 (1984).

\bibitem{GarciW1}
{A.L. Garcia, W. Wagner}, Phys. Rev. E {\bf 68}, 056703 (2003).

\bibitem{JacksZ1}
{B. Jackson, E. Zaremba}, Phys. Rev. {\bf A} {\bf 66}, 033606 (2002).

\bibitem{XuZZG1}
{Z. Xu, K. Zhou, P. Zhuang, C. Greiner}, Phys. Rev. Lett. {\bf 114}, 182301
  (2015).

\end{thebibliography}

\end{document}